\documentstyle[a4,12pt]{article}
\renewcommand{\theequation}{\thesection.\arabic{equation}}
\def\thefootnote{\fnsymbol{footnote}}
\renewcommand{\baselinestretch}{1.1}
\begin{document}
\parskip=5pt plus 1pt minus 1pt

\begin{flushright}
{\small\bf DPNU-97-18}\\
{March 1997}
\end{flushright}

\vspace{0.2cm}

\begin{center}
{\Large\bf Possible Signals of $D^0$-$\bar{D}^0$ Mixing and $CP$ Violation}
\footnote{Talk given at the Second International Conference on $B$ Physics
and $CP$ Violation, Honolulu, Hawaii, March 24 - 27, 1997 (to appear in the 
conference proceedings).}
\end{center}

\vspace{0.3cm}
\begin{center}
{\bf Zhi-zhong Xing} \footnote{Electronic address:
xing@eken.phys.nagoya-u.ac.jp}
\end{center}

\begin{center}
{\it Department of Physics, Nagoya University, Chikusa-ku, Nagoya 464-01, Japan}
\end{center}

\vspace{3.5cm}

\begin{abstract}
In view of the discovery potential associated with the future experiments of
high-luminosity
fixed target facilities, $B$-meson factories and $\tau$-charm factories,
we highlight some typical signals of $D^0$-$\bar{D}^0$ mixing and $CP$ violation
which are likely to show up in neutral $D$-meson decays to 
the semileptonic final states, the hadronic $CP$ eigenstates, the hadronic 
non-$CP$ eigenstates and the $CP$-forbidden states. Both time-dependent and
time-integrated measurements are discussed, and particular interest is paid
to $D^0/\bar{D}^0\rightarrow K_{S,L}+\pi^0$ and 
$D^0/\bar{D}^0\rightarrow K^{\pm}\pi^{\mp}$ transitions.
\end{abstract}

\newpage

\section{Introduction}

The study of $D^0$-$\bar{D}^0$ mixing and $CP$ violation in neutral $D$-meson
decays is not only complementary to our knowledge obtained and to be obtained from
$K^0$-$\bar{K}^0$ and $B^0$-$\bar{B}^0$ systems, but also important for exploring
underlying new physics that is out of reach of the standard model predictions. 
A large discovery potential associated with this topic is expected to exist in 
the future delicate experiments of high-luminosity fixed target facilities,
$B$-meson factories, and $\tau$-charm factories \cite{Proc}.

Without $CPT$ violation, the mass eigenstates of $D^0$ and $\bar{D}^0$ mesons
can be written as 
\begin{eqnarray}
| D_L \rangle \; = \; p |D^0\rangle ~ + ~ q |\bar{D}^0\rangle \; , \nonumber \\
| D_H \rangle \; = \; p |D^0\rangle ~ - ~ q |\bar{D}^0\rangle \; ,
\end{eqnarray}
where $p$ and $q$ are complex parameters determined by off-diagonal
elements of the $D^0$-$\bar{D}^0$ mixing Hamiltonian.
The rate of $D^0$-$\bar{D}^0$ mixing is commonly measured by two well-defined 
dimensionless quantities, 
\begin{equation}
x^{~}_D \; \equiv \; \frac{\Delta m}{\Gamma} \; , ~~~~~~~~
y^{~}_D \; \equiv \; \frac{\Delta \Gamma}{2 \Gamma} \; ,
\end{equation}
which correspond to the mass and width differences
of $D_H$ and $D_L$ (i.e., $\Delta m \equiv m^{~}_H - m^{~}_L$ and
$\Delta \Gamma \equiv \Gamma_L - \Gamma_H$).
The latest result from the Fermilab experiment E791 has set an upper bound on the rate 
of $D^0$-$\bar{D}^0$ mixing \cite{E791}:
\begin{equation}
r^{~}_D \; \equiv \; \frac{x^2_D + y^2_D}{2} \; < \; 5 \times 10^{-3} \; .
\end{equation}
In the standard model, the short-distance contribution to $D^0$-$\bar{D}^0$ mixing is via 
box diagrams and its magnitude is expected to be negligibly small.
But different approaches to the long-distance effects in $D^0$-$\bar{D}^0$ mixing,
which come mainly from the real intermediate states of $SU(3)$ multiplets, have given
dramatically different estimates for the magnitudes of $x^{~}_D$ and $y^{~}_D$
\cite{Golowich96}. If calculations based on the standard
model can reliably limit $x^{~}_D$ and $y^{~}_D$ to be well below $10^{-2}$, then
observation of $r^{~}_D$ at the level of $10^{-4}$ or so will imply the existence of
new physics \cite{Golowich97}. On the other hand, improved experimental knowledge of $r^{~}_D$, in
particular the relative magnitude of $x^{~}_D$ and $y^{~}_D$, can definitely clarify 
the ambiguities in current theoretical estimates and shed some light on both the dynamics
of $D^0$-$\bar{D}^0$ mixing and possible sources of new physics beyond
the standard model.

In principle, there may be three different types of $CP$-violating signals in
neutral $D$-meson transitions \cite{BigiSanda86,Xing97}:

(a) {\it $CP$ violation in $D^0$-$\bar{D}^0$ mixing}. This implies $|q/p|\neq 1$.
In practice, we have the following $CP$-violating observable:  
\begin{equation}
\Delta_D \; \equiv \; \frac{|p|^4 ~ - ~ |q|^4}{|p|^4 ~ + ~ |q|^4} \; .
\end{equation}
It is expected that the magnitude of $\Delta_D$ should be at most of the order 
$10^{-3}$ in the standard model. However, a reliable estimation of $\Delta_D$
suffers from large long-distance uncertainties.

(b) {\it $CP$ violation in direct decay}. For a decay mode $D^0\rightarrow f$ and
its $CP$-conjugate process $\bar{D}^0\rightarrow \bar{f}$, this implies 
\begin{equation}
|\langle \bar{f} |{\cal H}_{\rm eff}| \bar{D}^0 \rangle | 
\; \equiv \; \left | \sum_n \left [A_n e^{{\rm i} (\delta_n - \phi_n)} \right ] 
\right | ~~  \neq ~~
| \langle f |{\cal H}_{\rm eff}| D^0\rangle | \; \equiv \;
\left | \sum_n \left [ A_n e^{{\rm i} (\delta_n + \phi_n)} \right ] \right | \; ,
\end{equation}
where a parametrization of the decay amplitudes
with the weak ($\phi_n$) and strong ($\delta_n$) phases
is also given. We see that $n\geq 2$, $\phi_m - \phi_n \neq 0$ or $\pi$
and $\delta_m - \delta_n \neq 0$ or $\pi$ are necessary conditions for the 
above {\it direct} $CP$ violation.

(c) {\it $CP$ violation from the interplay of decay and mixing}. Let us define
two rephasing-invariant quantities
\begin{equation}
\lambda_f \; \equiv \; \frac{q}{p} \cdot \frac{\langle f|{\cal H}_{\rm eff}|\bar{D}^0\rangle}
{\langle f |{\cal H}_{\rm eff}|D^0\rangle} \; , ~~~~~~~~
\bar{\lambda}_{\bar f} \; \equiv \; \frac{p}{q} \cdot \frac{\langle \bar{f}|{\cal H}_{\rm eff}|
D^0\rangle }{\langle \bar{f} |{\cal H}_{\rm eff}|\bar{D}^0\rangle } \; ,
\end{equation}
where the hadronic states $f$ and $\bar{f}$ are common to the decay of $D^0$ 
(or $\bar{D}^0$). Even in the assumption of $|q/p|=1$, {\it indirect} $CP$
violation can appear if 
\begin{equation}
{\rm Im} \lambda_f ~ - ~ {\rm Im}\bar{\lambda}_{\bar f} \; \neq \; 0 \; .
\end{equation}
Provided $f$ is a $CP$ eigenstate (i.e., $|\bar{f}\rangle = \pm |f\rangle$) and the 
decay is dominated by a single weak phase, then we have $\bar{\lambda}_{\bar f}
= \lambda^*_f$. 

$CP$ violation at the percent level has not been observed in experiments \cite{CLEO-E687}. 
But signals of $O (10^{-3})$ are expected in some
neutral $D$ decays within the standard model, and those of $O(10^{-2})$
cannot be ruled out in some channels beyond the standard model. 

Subsequently we shall highlight some typical signals of $D^0$-$\bar{D}^0$ mixing
and $CP$ violation which are likely to show up in weak decays of neutral $D$ mesons. 
A systematic and comprehensive study of this topic can be found in 
Ref. \cite{Xing97} and references therein.

\section{Typical signals of $D^0$-$\bar{D}^0$ mixing}
\setcounter{equation}{0}

For simplicity and instruction, we assume $\Delta_D =0$ in the discussion of 
$D^0$-$\bar{D}^0$ mixing effects. This assumption should be safe both within
and beyond the standard model, and it can be tested by detecting $CP$ violation
in the semileptonic decays of $D^0$ and $\bar{D}^0$ mesons.

\begin{center}
{\large\bf A. ~ Time-integrated measurements}
\end{center}

For fixed target experiments or $e^{+}e^{-}$ collisions at the $\Upsilon(4S)$ resonance,
the produced $D^0$ and $\bar{D}^0$ mesons are incoherent. Knowledge of $D^0$-$\bar{D}^0$
mixing is expected to come from ratios of the wrong-sign to right-sign events of
semileptonic $D$ decays:
\begin{equation}
\frac{{\cal R}(D^0_{\rm phys}\rightarrow K^+ l^- \bar{\nu}^{~}_l)}
{{\cal R}(D^0_{\rm phys}\rightarrow K^- l^+ \nu^{~}_l)} \; 
\approx \; \frac{{\cal R}(\bar{D}^0_{\rm phys}\rightarrow K^- l^+ \nu^{~}_l)}
{{\cal R}(\bar{D}^0_{\rm phys}\rightarrow K^+ l^- \bar{\nu}^{~}_l)} \; \approx \; r^{~}_D 
\end{equation}
in the assumption made above. The Fermilab experiment E791 gives $r^{~}_D <
0.5\%$ at the $90\%$ confidence level \cite{E791}, the best model-independent
limit on $D^0$-$\bar{D}^0$ mixing today.

For a $\tau$-charm factory running on the $\psi (3.77)$ resonance, coherent 
$D^0\bar{D}^0$ events with odd $C$-parity can be produced. If $e^+e^-$
collisions take place at the $\psi(4.16)$ resonance, coherent 
$D^0\bar{D}^0$ events will be produced through $\psi(4.16) \rightarrow \gamma 
(D^0\bar{D}^0)_{C-\rm even}$ or $\psi(4.16)\rightarrow \pi^0 (D^0\bar{D}^0)_{C-\rm odd}$.
Three types of joint decay modes are interesting for measuring $D^0$-$\bar{D}^0$ mixing:
\begin{eqnarray}
(D^0_{\rm phys} \bar{D}^0_{\rm phys})_C & \longrightarrow & 
(l^{\pm} X^{\mp})_D ~ (l^{\pm} X^{\mp})_{\bar D} \; , \nonumber \\
& \longrightarrow & (K^{\pm} \pi^{\mp})_D ~ (l^{\mp} X^{\pm})_{\bar D} 
\; , \nonumber \\
& \longrightarrow & 
(K^{\pm} \pi^{\mp})_D ~ (K^{\pm} \pi^{\mp})_{\bar D} \; ,
\end{eqnarray}
where we have used the notations $X^+ \equiv K^+ \bar{\nu}^{~}_l$ and $X^- \equiv K^- \nu^{~}_l$. 
Note that $D^0\rightarrow K^+ \pi^-$ is a doubly Cabibbo-suppressed
decay (DCSD). This effect is usually measured by the following ratio
of decay rates:
\begin{equation}
R_{\rm\scriptstyle DCSD} \; \equiv \; \left | \frac{\langle K^+\pi^-|{\cal H}_{\rm eff}
|D^0\rangle}{\langle K^-\pi^+|{\cal H}_{\rm eff}|D^0\rangle} \right |^2 \; .
\end{equation}
In the assumption of $r^{~}_D =0$, $R_{\rm\scriptstyle DCSD} \approx 0.77\%$
and $0.68\%$ are respectively obtained by CLEO II and Fermilab E791 experiments
\cite{CLEO-E791}.
For our present purpose, we list the possible signals of $D^0$-$\bar{D}^0$
mixing associated with the joint decays (2.2) in Table 1, where $|q/p|=1$ has
been used and the interference terms $T^{\pm}_{\rm int}$ are given by
\begin{eqnarray}
T^+_{\rm int} & = & \sqrt{R_{\rm\scriptstyle DCSD}} ~ \left [ y^{~}_D 
\cos (\delta_{K\pi} - \phi_D) ~ - ~ x^{~}_D \sin (\delta_{K\pi} - \phi_D) 
\right ] \; , \nonumber \\
T^-_{\rm int} & = & \sqrt{R_{\rm\scriptstyle DCSD}} ~ \left [ y^{~}_D
\cos (\delta_{K\pi} + \phi_D) ~ - ~ x^{~}_D \sin (\delta_{K\pi} + \phi_D)
\right ] \; 
\end{eqnarray}
with $\phi_D \equiv \arg (q/p)$ and $\delta_{K\pi} \equiv \arg (\langle
K^+\pi^-|{\cal H}_{\rm eff}|D^0\rangle / \langle K^-\pi^+ |{\cal H}_{\rm eff}|
D^0\rangle )$. Here we have assumed $\delta_{K\pi}$ to be a pure strong
phase shift by neglecting the tiny weak phase ($\sim 10^{-4}$) from the 
Cabibbo-Kobayashi-Maskawa (CKM) matrix elements.
\begin{center}
{Table 1: $D^0$-$\bar{D}^0$ mixing and DCSD effects in three types of 
coherent $D^0\bar{D}^0$ decays \cite{Xing97,Xing96}.} 
\begin{tabular}{ccc} \\ \hline\hline \\ 
~~~~~ Observable ~~~~~	& ~~~~~ Signal ~ ($C$-odd) ~~~~~	
& ~~~~~ Signal ~ ($C$-even) ~~~~~ \\ \\ \hline \\
$\displaystyle \frac{{\cal R} (l^{\pm} X^{\mp}; l^{\pm} X^{\mp})_C}
{{\cal R} (l^{\pm} X^{\mp}; l^{\mp} X^{\pm})_C}$ 	
& $r^{~}_D$	& $ 3 r^{~}_D$ 	\\ \\ \hline \\
$\displaystyle \frac{{\cal R} (K^{\pm} \pi^{\mp}; l^{\mp} X^{\pm})_C}
{{\cal R} (K^{\pm} \pi^{\mp}; l^{\pm} X^{\mp})_C}$
& $R_{\rm\scriptstyle DCSD} ~ + ~ r^{~}_D$	& $R_{\rm\scriptstyle DCSD} ~ 
+ ~ 3 r^{~}_D ~ + ~ 2 T^{\pm}_{\rm int}$ \\ \\ \hline \\
$\displaystyle \frac{{\cal R} (K^{\pm} \pi^{\mp}; K^{\pm} \pi^{\mp})_C}
{{\cal R} (K^{\pm} \pi^{\mp}; K^{\mp} \pi^{\pm})_C}$
& $r^{~}_D$	& $4 R_{\rm\scriptstyle DCSD} ~ + ~ 3 r^{~}_D ~ + ~ 
4 T^{\pm}_{\rm int}$ \\ \\ \hline \hline
\end{tabular}
\end{center}

We observe that $r^{~}_{\rm D}$, $R_{\rm\scriptstyle DCSD}$ and
$T^{\pm}_{\rm int}$ can all be determined from measurements of the
above joint decays, only if the size of $r^{~}_D$ is comparable 
with that of $R_{\rm\scriptstyle DCSD}$. In particular, the
information about $T^{\pm}_{\rm int}$ is useful to give some hints 
on the relative size of $x^{~}_D$ and $y^{~}_D$ as well as the
$CP$-violating phase $\phi_D$. In the case that $D^0$-$\bar{D}^0$
mixing is negligibly small, we are able to isolate the DCSD rate
$R_{\rm\scriptstyle DCSD}$ solely on the $\psi (3.77)$ resonance.

\begin{center}
{\large\bf B. ~ Time-dependent measurements}
\end{center}

Let us illustrate two examples about the time-dependent measurement of 
$D^0$-$\bar{D}^0$ mixing. The first is associated with the well-known
DCSDs $D^0\rightarrow K^+\pi^-$ and $\bar{D}^0\rightarrow K^-\pi^+$
\cite{Pakvasa96Wol95Nir95}. The time-dependent decay rates of these two modes, 
in comparison with
the rates of their Cabibbo-allowed counterparts, read as follows:
\begin{eqnarray}
\frac{{\cal R} [D^0 (t) \rightarrow K^+\pi^-]}{{\cal R}
[D^0 (t) \rightarrow K^-\pi^+]} & = & R_{\rm\scriptstyle DCSD}
~ + ~ T^+_{\rm int} \left (\Gamma t \right ) ~ + ~ 
\frac{r^{~}_D}{2} \left (\Gamma t \right )^2 \; ,
\nonumber \\
\frac{{\cal R} [\bar{D}^0 (t) \rightarrow K^-\pi^+]}{{\cal R}
[\bar{D}^0 (t) \rightarrow K^+\pi^-]} & = & R_{\rm\scriptstyle DCSD}
~ + ~ T^-_{\rm int} \left (\Gamma t \right ) ~ + ~ 
\frac{r^{~}_D}{2} \left (\Gamma t \right )^2 \; ,
\end{eqnarray}
where $\Delta_D =0$ has been assumed. Just for the purpose of illustration,
we take $x^{~}_D = 0.05$, $y^{~}_D =0$, $\delta_{K\pi} =0$, $\phi_D = \pi/2$
and $R_{\rm\scriptstyle DCSD} = 0.7\%$ to plot changes of the above two observables
with proper time $t$ in Fig. 1. We see that nonvanishing $D^0$-$\bar{D}^0$ mixing can
give rise to detectable time-evolution behaviors in 
${\cal R} [D^0(t)\rightarrow K^+\pi^-] / {\cal R} [D^0(t)\rightarrow K^-\pi^+]$
and ${\cal R} [\bar{D}^0(t)\rightarrow K^-\pi^+] / {\cal R} [\bar{D}^0(t)
\rightarrow K^+\pi^-]$, while the difference between these two ratios 
comes from $CP$ violation hidden in the interference terms $T^{\pm}_{\rm int}$.
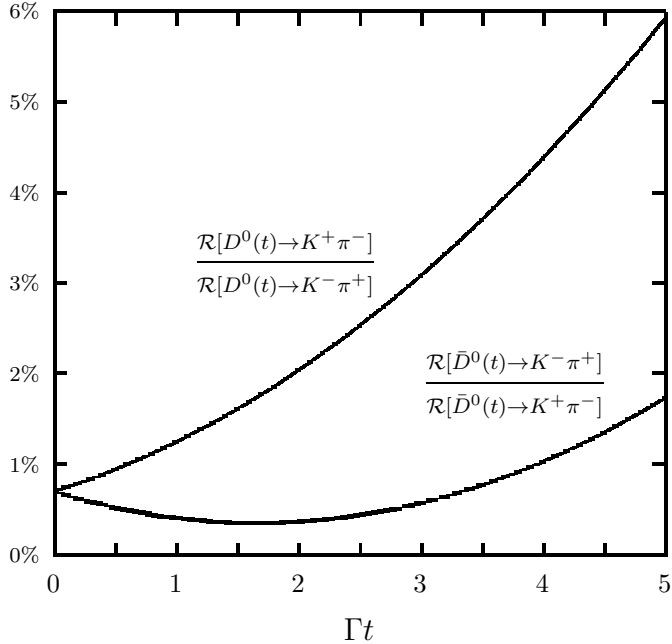
\begin{figure}
\setlength{\unitlength}{0.240900pt}
\ifx\plotpoint\undefined\newsavebox{\plotpoint}\fi
\sbox{\plotpoint}{\rule[-0.500pt]{1.000pt}{1.000pt}}%
\begin{picture}(1200,990)(-350,0)
\font\gnuplot=cmr10 at 10pt
\gnuplot
\sbox{\plotpoint}{\rule[-0.500pt]{1.000pt}{1.000pt}}%
\put(176.0,113.0){\rule[-0.500pt]{231.264pt}{1.000pt}}
\put(176.0,113.0){\rule[-0.500pt]{1.000pt}{205.729pt}}
\put(176.0,113.0){\rule[-0.500pt]{4.818pt}{1.000pt}}
\put(154,113){\makebox(0,0)[r]{$\scriptstyle 0\%$}}
\put(1116.0,113.0){\rule[-0.500pt]{4.818pt}{1.000pt}}
\put(176.0,255.0){\rule[-0.500pt]{4.818pt}{1.000pt}}
\put(154,255){\makebox(0,0)[r]{$\scriptstyle 1\%$}}
\put(1116.0,255.0){\rule[-0.500pt]{4.818pt}{1.000pt}}
\put(176.0,398.0){\rule[-0.500pt]{4.818pt}{1.000pt}}
\put(154,398){\makebox(0,0)[r]{$\scriptstyle 2\%$}}
\put(1116.0,398.0){\rule[-0.500pt]{4.818pt}{1.000pt}}
\put(176.0,540.0){\rule[-0.500pt]{4.818pt}{1.000pt}}
\put(154,540){\makebox(0,0)[r]{$\scriptstyle 3\%$}}
\put(1116.0,540.0){\rule[-0.500pt]{4.818pt}{1.000pt}}
\put(176.0,682.0){\rule[-0.500pt]{4.818pt}{1.000pt}}
\put(154,682){\makebox(0,0)[r]{$\scriptstyle 4\%$}}
\put(1116.0,682.0){\rule[-0.500pt]{4.818pt}{1.000pt}}
\put(176.0,825.0){\rule[-0.500pt]{4.818pt}{1.000pt}}
\put(154,825){\makebox(0,0)[r]{$\scriptstyle 5\%$}}
\put(1116.0,825.0){\rule[-0.500pt]{4.818pt}{1.000pt}}
\put(176.0,967.0){\rule[-0.500pt]{4.818pt}{1.000pt}}
\put(154,967){\makebox(0,0)[r]{$\scriptstyle 6\%$}}
\put(1116.0,967.0){\rule[-0.500pt]{4.818pt}{1.000pt}}
\put(176.0,113.0){\rule[-0.500pt]{1.000pt}{4.818pt}}
\put(176,68){\makebox(0,0){0}}
\put(176.0,947.0){\rule[-0.500pt]{1.000pt}{4.818pt}}
\put(272.0,113.0){\rule[-0.500pt]{1.000pt}{4.818pt}}
\put(272.0,947.0){\rule[-0.500pt]{1.000pt}{4.818pt}}
\put(368.0,113.0){\rule[-0.500pt]{1.000pt}{4.818pt}}
\put(368,68){\makebox(0,0){1}}
\put(368.0,947.0){\rule[-0.500pt]{1.000pt}{4.818pt}}
\put(464.0,113.0){\rule[-0.500pt]{1.000pt}{4.818pt}}
\put(464.0,947.0){\rule[-0.500pt]{1.000pt}{4.818pt}}
\put(560.0,113.0){\rule[-0.500pt]{1.000pt}{4.818pt}}
\put(560,68){\makebox(0,0){2}}
\put(560.0,947.0){\rule[-0.500pt]{1.000pt}{4.818pt}}
\put(656.0,113.0){\rule[-0.500pt]{1.000pt}{4.818pt}}
\put(656.0,947.0){\rule[-0.500pt]{1.000pt}{4.818pt}}
\put(752.0,113.0){\rule[-0.500pt]{1.000pt}{4.818pt}}
\put(752,68){\makebox(0,0){3}}
\put(752.0,947.0){\rule[-0.500pt]{1.000pt}{4.818pt}}
\put(848.0,113.0){\rule[-0.500pt]{1.000pt}{4.818pt}}
\put(848.0,947.0){\rule[-0.500pt]{1.000pt}{4.818pt}}
\put(944.0,113.0){\rule[-0.500pt]{1.000pt}{4.818pt}}
\put(944,68){\makebox(0,0){4}}
\put(944.0,947.0){\rule[-0.500pt]{1.000pt}{4.818pt}}
\put(1040.0,113.0){\rule[-0.500pt]{1.000pt}{4.818pt}}
\put(1040.0,947.0){\rule[-0.500pt]{1.000pt}{4.818pt}}
\put(1136.0,113.0){\rule[-0.500pt]{1.000pt}{4.818pt}}
\put(1136,68){\makebox(0,0){5}}
\put(1136.0,947.0){\rule[-0.500pt]{1.000pt}{4.818pt}}
\put(176.0,113.0){\rule[-0.500pt]{231.264pt}{1.000pt}}
\put(1136.0,113.0){\rule[-0.500pt]{1.000pt}{205.729pt}}
\put(176.0,967.0){\rule[-0.500pt]{231.264pt}{1.000pt}}
\put(656,-3){\makebox(0,0){$\Gamma t$}}
\put(540,570){\makebox(0,0){$\displaystyle\frac{\scriptstyle {\cal R} [D^0 (t)
\rightarrow K^+\pi^-]}{\scriptstyle {\cal R} [D^0 (t) \rightarrow
K^-\pi^+]}$}}
\put(900,380){\makebox(0,0){$\displaystyle\frac{\scriptstyle {\cal R} [\bar{D}^0 (t)
\rightarrow K^-\pi^+]}{\scriptstyle {\cal R} [\bar{D}^0 (t) \rightarrow
K^+\pi^-]}$}}

\put(176.0,113.0){\rule[-0.500pt]{1.000pt}{205.729pt}}
\put(176,213){\usebox{\plotpoint}}
\multiput(176.00,214.84)(1.606,0.462){4}{\rule{3.417pt}{0.111pt}}
\multiput(176.00,210.92)(11.909,6.000){2}{\rule{1.708pt}{1.000pt}}
\multiput(195.00,220.84)(1.606,0.462){4}{\rule{3.417pt}{0.111pt}}
\multiput(195.00,216.92)(11.909,6.000){2}{\rule{1.708pt}{1.000pt}}
\multiput(214.00,226.84)(1.420,0.475){6}{\rule{3.107pt}{0.114pt}}
\multiput(214.00,222.92)(13.551,7.000){2}{\rule{1.554pt}{1.000pt}}
\multiput(234.00,233.84)(1.339,0.475){6}{\rule{2.964pt}{0.114pt}}
\multiput(234.00,229.92)(12.847,7.000){2}{\rule{1.482pt}{1.000pt}}
\multiput(253.00,240.83)(1.158,0.481){8}{\rule{2.625pt}{0.116pt}}
\multiput(253.00,236.92)(13.552,8.000){2}{\rule{1.313pt}{1.000pt}}
\multiput(272.00,248.83)(1.158,0.481){8}{\rule{2.625pt}{0.116pt}}
\multiput(272.00,244.92)(13.552,8.000){2}{\rule{1.313pt}{1.000pt}}
\multiput(291.00,256.83)(1.158,0.481){8}{\rule{2.625pt}{0.116pt}}
\multiput(291.00,252.92)(13.552,8.000){2}{\rule{1.313pt}{1.000pt}}
\multiput(310.00,264.83)(1.082,0.485){10}{\rule{2.472pt}{0.117pt}}
\multiput(310.00,260.92)(14.869,9.000){2}{\rule{1.236pt}{1.000pt}}
\multiput(330.00,273.83)(1.022,0.485){10}{\rule{2.361pt}{0.117pt}}
\multiput(330.00,269.92)(14.099,9.000){2}{\rule{1.181pt}{1.000pt}}
\multiput(349.00,282.83)(1.022,0.485){10}{\rule{2.361pt}{0.117pt}}
\multiput(349.00,278.92)(14.099,9.000){2}{\rule{1.181pt}{1.000pt}}
\multiput(368.00,291.83)(0.916,0.487){12}{\rule{2.150pt}{0.117pt}}
\multiput(368.00,287.92)(14.538,10.000){2}{\rule{1.075pt}{1.000pt}}
\multiput(387.00,301.83)(0.916,0.487){12}{\rule{2.150pt}{0.117pt}}
\multiput(387.00,297.92)(14.538,10.000){2}{\rule{1.075pt}{1.000pt}}
\multiput(406.00,311.83)(0.969,0.487){12}{\rule{2.250pt}{0.117pt}}
\multiput(406.00,307.92)(15.330,10.000){2}{\rule{1.125pt}{1.000pt}}
\multiput(426.00,321.83)(0.830,0.489){14}{\rule{1.977pt}{0.118pt}}
\multiput(426.00,317.92)(14.896,11.000){2}{\rule{0.989pt}{1.000pt}}
\multiput(445.00,332.83)(0.830,0.489){14}{\rule{1.977pt}{0.118pt}}
\multiput(445.00,328.92)(14.896,11.000){2}{\rule{0.989pt}{1.000pt}}
\multiput(464.00,343.83)(0.830,0.489){14}{\rule{1.977pt}{0.118pt}}
\multiput(464.00,339.92)(14.896,11.000){2}{\rule{0.989pt}{1.000pt}}
\multiput(483.00,354.83)(0.759,0.491){16}{\rule{1.833pt}{0.118pt}}
\multiput(483.00,350.92)(15.195,12.000){2}{\rule{0.917pt}{1.000pt}}
\multiput(502.00,366.83)(0.803,0.491){16}{\rule{1.917pt}{0.118pt}}
\multiput(502.00,362.92)(16.022,12.000){2}{\rule{0.958pt}{1.000pt}}
\multiput(522.00,378.83)(0.700,0.492){18}{\rule{1.712pt}{0.118pt}}
\multiput(522.00,374.92)(15.448,13.000){2}{\rule{0.856pt}{1.000pt}}
\multiput(541.00,391.83)(0.700,0.492){18}{\rule{1.712pt}{0.118pt}}
\multiput(541.00,387.92)(15.448,13.000){2}{\rule{0.856pt}{1.000pt}}
\multiput(560.00,404.83)(0.700,0.492){18}{\rule{1.712pt}{0.118pt}}
\multiput(560.00,400.92)(15.448,13.000){2}{\rule{0.856pt}{1.000pt}}
\multiput(579.00,417.83)(0.649,0.492){20}{\rule{1.607pt}{0.119pt}}
\multiput(579.00,413.92)(15.664,14.000){2}{\rule{0.804pt}{1.000pt}}
\multiput(598.00,431.83)(0.686,0.492){20}{\rule{1.679pt}{0.119pt}}
\multiput(598.00,427.92)(16.516,14.000){2}{\rule{0.839pt}{1.000pt}}
\multiput(618.00,445.83)(0.649,0.492){20}{\rule{1.607pt}{0.119pt}}
\multiput(618.00,441.92)(15.664,14.000){2}{\rule{0.804pt}{1.000pt}}
\multiput(637.00,459.83)(0.605,0.493){22}{\rule{1.517pt}{0.119pt}}
\multiput(637.00,455.92)(15.852,15.000){2}{\rule{0.758pt}{1.000pt}}
\multiput(656.00,474.83)(0.605,0.493){22}{\rule{1.517pt}{0.119pt}}
\multiput(656.00,470.92)(15.852,15.000){2}{\rule{0.758pt}{1.000pt}}
\multiput(675.00,489.83)(0.605,0.493){22}{\rule{1.517pt}{0.119pt}}
\multiput(675.00,485.92)(15.852,15.000){2}{\rule{0.758pt}{1.000pt}}
\multiput(694.00,504.83)(0.599,0.494){24}{\rule{1.500pt}{0.119pt}}
\multiput(694.00,500.92)(16.887,16.000){2}{\rule{0.750pt}{1.000pt}}
\multiput(714.00,520.83)(0.567,0.494){24}{\rule{1.438pt}{0.119pt}}
\multiput(714.00,516.92)(16.016,16.000){2}{\rule{0.719pt}{1.000pt}}
\multiput(733.00,536.83)(0.567,0.494){24}{\rule{1.438pt}{0.119pt}}
\multiput(733.00,532.92)(16.016,16.000){2}{\rule{0.719pt}{1.000pt}}
\multiput(752.00,552.83)(0.533,0.494){26}{\rule{1.368pt}{0.119pt}}
\multiput(752.00,548.92)(16.161,17.000){2}{\rule{0.684pt}{1.000pt}}
\multiput(771.00,569.83)(0.533,0.494){26}{\rule{1.368pt}{0.119pt}}
\multiput(771.00,565.92)(16.161,17.000){2}{\rule{0.684pt}{1.000pt}}
\multiput(790.00,586.83)(0.531,0.495){28}{\rule{1.361pt}{0.119pt}}
\multiput(790.00,582.92)(17.175,18.000){2}{\rule{0.681pt}{1.000pt}}
\multiput(810.00,604.83)(0.503,0.495){28}{\rule{1.306pt}{0.119pt}}
\multiput(810.00,600.92)(16.290,18.000){2}{\rule{0.653pt}{1.000pt}}
\multiput(829.00,622.83)(0.503,0.495){28}{\rule{1.306pt}{0.119pt}}
\multiput(829.00,618.92)(16.290,18.000){2}{\rule{0.653pt}{1.000pt}}
\multiput(848.00,640.83)(0.476,0.495){30}{\rule{1.250pt}{0.119pt}}
\multiput(848.00,636.92)(16.406,19.000){2}{\rule{0.625pt}{1.000pt}}
\multiput(867.00,659.83)(0.476,0.495){30}{\rule{1.250pt}{0.119pt}}
\multiput(867.00,655.92)(16.406,19.000){2}{\rule{0.625pt}{1.000pt}}
\multiput(886.00,678.83)(0.503,0.495){30}{\rule{1.303pt}{0.119pt}}
\multiput(886.00,674.92)(17.296,19.000){2}{\rule{0.651pt}{1.000pt}}
\multiput(906.00,697.83)(0.476,0.495){30}{\rule{1.250pt}{0.119pt}}
\multiput(906.00,693.92)(16.406,19.000){2}{\rule{0.625pt}{1.000pt}}
\multiput(926.83,715.00)(0.495,0.503){30}{\rule{0.119pt}{1.303pt}}
\multiput(922.92,715.00)(19.000,17.296){2}{\rule{1.000pt}{0.651pt}}
\multiput(945.83,735.00)(0.495,0.530){30}{\rule{0.119pt}{1.355pt}}
\multiput(941.92,735.00)(19.000,18.187){2}{\rule{1.000pt}{0.678pt}}
\multiput(964.83,756.00)(0.495,0.530){30}{\rule{0.119pt}{1.355pt}}
\multiput(960.92,756.00)(19.000,18.187){2}{\rule{1.000pt}{0.678pt}}
\multiput(983.83,777.00)(0.495,0.503){32}{\rule{0.119pt}{1.300pt}}
\multiput(979.92,777.00)(20.000,18.302){2}{\rule{1.000pt}{0.650pt}}
\multiput(1003.83,798.00)(0.495,0.530){30}{\rule{0.119pt}{1.355pt}}
\multiput(999.92,798.00)(19.000,18.187){2}{\rule{1.000pt}{0.678pt}}
\multiput(1022.83,819.00)(0.495,0.557){30}{\rule{0.119pt}{1.408pt}}
\multiput(1018.92,819.00)(19.000,19.078){2}{\rule{1.000pt}{0.704pt}}
\multiput(1041.83,841.00)(0.495,0.557){30}{\rule{0.119pt}{1.408pt}}
\multiput(1037.92,841.00)(19.000,19.078){2}{\rule{1.000pt}{0.704pt}}
\multiput(1060.83,863.00)(0.495,0.557){30}{\rule{0.119pt}{1.408pt}}
\multiput(1056.92,863.00)(19.000,19.078){2}{\rule{1.000pt}{0.704pt}}
\multiput(1079.83,885.00)(0.495,0.554){32}{\rule{0.119pt}{1.400pt}}
\multiput(1075.92,885.00)(20.000,20.094){2}{\rule{1.000pt}{0.700pt}}
\multiput(1099.83,908.00)(0.495,0.611){30}{\rule{0.119pt}{1.513pt}}
\multiput(1095.92,908.00)(19.000,20.859){2}{\rule{1.000pt}{0.757pt}}
\multiput(1118.83,932.00)(0.495,0.584){30}{\rule{0.119pt}{1.461pt}}
\multiput(1114.92,932.00)(19.000,19.969){2}{\rule{1.000pt}{0.730pt}}
\put(176,213){\usebox{\plotpoint}}
\multiput(176.00,210.69)(1.606,-0.462){4}{\rule{3.417pt}{0.111pt}}
\multiput(176.00,210.92)(11.909,-6.000){2}{\rule{1.708pt}{1.000pt}}
\multiput(195.00,204.69)(1.606,-0.462){4}{\rule{3.417pt}{0.111pt}}
\multiput(195.00,204.92)(11.909,-6.000){2}{\rule{1.708pt}{1.000pt}}
\multiput(214.00,198.71)(2.358,-0.424){2}{\rule{4.250pt}{0.102pt}}
\multiput(214.00,198.92)(11.179,-5.000){2}{\rule{2.125pt}{1.000pt}}
\put(234,191.92){\rule{4.577pt}{1.000pt}}
\multiput(234.00,193.92)(9.500,-4.000){2}{\rule{2.289pt}{1.000pt}}
\multiput(253.00,189.71)(2.188,-0.424){2}{\rule{4.050pt}{0.102pt}}
\multiput(253.00,189.92)(10.594,-5.000){2}{\rule{2.025pt}{1.000pt}}
\put(272,182.92){\rule{4.577pt}{1.000pt}}
\multiput(272.00,184.92)(9.500,-4.000){2}{\rule{2.289pt}{1.000pt}}
\put(291,179.42){\rule{4.577pt}{1.000pt}}
\multiput(291.00,180.92)(9.500,-3.000){2}{\rule{2.289pt}{1.000pt}}
\put(310,175.92){\rule{4.818pt}{1.000pt}}
\multiput(310.00,177.92)(10.000,-4.000){2}{\rule{2.409pt}{1.000pt}}
\put(330,172.42){\rule{4.577pt}{1.000pt}}
\multiput(330.00,173.92)(9.500,-3.000){2}{\rule{2.289pt}{1.000pt}}
\put(349,169.92){\rule{4.577pt}{1.000pt}}
\multiput(349.00,170.92)(9.500,-2.000){2}{\rule{2.289pt}{1.000pt}}
\put(368,167.92){\rule{4.577pt}{1.000pt}}
\multiput(368.00,168.92)(9.500,-2.000){2}{\rule{2.289pt}{1.000pt}}
\put(387,165.92){\rule{4.577pt}{1.000pt}}
\multiput(387.00,166.92)(9.500,-2.000){2}{\rule{2.289pt}{1.000pt}}
\put(406,163.92){\rule{4.818pt}{1.000pt}}
\multiput(406.00,164.92)(10.000,-2.000){2}{\rule{2.409pt}{1.000pt}}
\put(426,162.42){\rule{4.577pt}{1.000pt}}
\multiput(426.00,162.92)(9.500,-1.000){2}{\rule{2.289pt}{1.000pt}}
\put(445,161.42){\rule{4.577pt}{1.000pt}}
\multiput(445.00,161.92)(9.500,-1.000){2}{\rule{2.289pt}{1.000pt}}
\put(522,161.42){\rule{4.577pt}{1.000pt}}
\multiput(522.00,160.92)(9.500,1.000){2}{\rule{2.289pt}{1.000pt}}
\put(541,162.42){\rule{4.577pt}{1.000pt}}
\multiput(541.00,161.92)(9.500,1.000){2}{\rule{2.289pt}{1.000pt}}
\put(560,163.42){\rule{4.577pt}{1.000pt}}
\multiput(560.00,162.92)(9.500,1.000){2}{\rule{2.289pt}{1.000pt}}
\put(579,164.92){\rule{4.577pt}{1.000pt}}
\multiput(579.00,163.92)(9.500,2.000){2}{\rule{2.289pt}{1.000pt}}
\put(598,166.92){\rule{4.818pt}{1.000pt}}
\multiput(598.00,165.92)(10.000,2.000){2}{\rule{2.409pt}{1.000pt}}
\put(618,168.92){\rule{4.577pt}{1.000pt}}
\multiput(618.00,167.92)(9.500,2.000){2}{\rule{2.289pt}{1.000pt}}
\put(637,171.42){\rule{4.577pt}{1.000pt}}
\multiput(637.00,169.92)(9.500,3.000){2}{\rule{2.289pt}{1.000pt}}
\put(656,174.42){\rule{4.577pt}{1.000pt}}
\multiput(656.00,172.92)(9.500,3.000){2}{\rule{2.289pt}{1.000pt}}
\put(675,177.92){\rule{4.577pt}{1.000pt}}
\multiput(675.00,175.92)(9.500,4.000){2}{\rule{2.289pt}{1.000pt}}
\put(694,181.42){\rule{4.818pt}{1.000pt}}
\multiput(694.00,179.92)(10.000,3.000){2}{\rule{2.409pt}{1.000pt}}
\multiput(714.00,186.86)(2.188,0.424){2}{\rule{4.050pt}{0.102pt}}
\multiput(714.00,182.92)(10.594,5.000){2}{\rule{2.025pt}{1.000pt}}
\put(733,189.92){\rule{4.577pt}{1.000pt}}
\multiput(733.00,187.92)(9.500,4.000){2}{\rule{2.289pt}{1.000pt}}
\multiput(752.00,195.86)(2.188,0.424){2}{\rule{4.050pt}{0.102pt}}
\multiput(752.00,191.92)(10.594,5.000){2}{\rule{2.025pt}{1.000pt}}
\multiput(771.00,200.86)(2.188,0.424){2}{\rule{4.050pt}{0.102pt}}
\multiput(771.00,196.92)(10.594,5.000){2}{\rule{2.025pt}{1.000pt}}
\multiput(790.00,205.84)(1.708,0.462){4}{\rule{3.583pt}{0.111pt}}
\multiput(790.00,201.92)(12.563,6.000){2}{\rule{1.792pt}{1.000pt}}
\multiput(810.00,211.84)(1.606,0.462){4}{\rule{3.417pt}{0.111pt}}
\multiput(810.00,207.92)(11.909,6.000){2}{\rule{1.708pt}{1.000pt}}
\multiput(829.00,217.84)(1.606,0.462){4}{\rule{3.417pt}{0.111pt}}
\multiput(829.00,213.92)(11.909,6.000){2}{\rule{1.708pt}{1.000pt}}
\multiput(848.00,223.84)(1.339,0.475){6}{\rule{2.964pt}{0.114pt}}
\multiput(848.00,219.92)(12.847,7.000){2}{\rule{1.482pt}{1.000pt}}
\multiput(867.00,230.84)(1.339,0.475){6}{\rule{2.964pt}{0.114pt}}
\multiput(867.00,226.92)(12.847,7.000){2}{\rule{1.482pt}{1.000pt}}
\multiput(886.00,237.84)(1.420,0.475){6}{\rule{3.107pt}{0.114pt}}
\multiput(886.00,233.92)(13.551,7.000){2}{\rule{1.554pt}{1.000pt}}
\multiput(906.00,244.83)(1.158,0.481){8}{\rule{2.625pt}{0.116pt}}
\multiput(906.00,240.92)(13.552,8.000){2}{\rule{1.313pt}{1.000pt}}
\multiput(925.00,252.83)(1.158,0.481){8}{\rule{2.625pt}{0.116pt}}
\multiput(925.00,248.92)(13.552,8.000){2}{\rule{1.313pt}{1.000pt}}
\multiput(944.00,260.83)(1.022,0.485){10}{\rule{2.361pt}{0.117pt}}
\multiput(944.00,256.92)(14.099,9.000){2}{\rule{1.181pt}{1.000pt}}
\multiput(963.00,269.83)(1.158,0.481){8}{\rule{2.625pt}{0.116pt}}
\multiput(963.00,265.92)(13.552,8.000){2}{\rule{1.313pt}{1.000pt}}
\multiput(982.00,277.83)(0.969,0.487){12}{\rule{2.250pt}{0.117pt}}
\multiput(982.00,273.92)(15.330,10.000){2}{\rule{1.125pt}{1.000pt}}
\multiput(1002.00,287.83)(1.022,0.485){10}{\rule{2.361pt}{0.117pt}}
\multiput(1002.00,283.92)(14.099,9.000){2}{\rule{1.181pt}{1.000pt}}
\multiput(1021.00,296.83)(0.916,0.487){12}{\rule{2.150pt}{0.117pt}}
\multiput(1021.00,292.92)(14.538,10.000){2}{\rule{1.075pt}{1.000pt}}
\multiput(1040.00,306.83)(0.916,0.487){12}{\rule{2.150pt}{0.117pt}}
\multiput(1040.00,302.92)(14.538,10.000){2}{\rule{1.075pt}{1.000pt}}
\multiput(1059.00,316.83)(0.830,0.489){14}{\rule{1.977pt}{0.118pt}}
\multiput(1059.00,312.92)(14.896,11.000){2}{\rule{0.989pt}{1.000pt}}
\multiput(1078.00,327.83)(0.878,0.489){14}{\rule{2.068pt}{0.118pt}}
\multiput(1078.00,323.92)(15.707,11.000){2}{\rule{1.034pt}{1.000pt}}
\multiput(1098.00,338.83)(0.830,0.489){14}{\rule{1.977pt}{0.118pt}}
\multiput(1098.00,334.92)(14.896,11.000){2}{\rule{0.989pt}{1.000pt}}
\multiput(1117.00,349.83)(0.759,0.491){16}{\rule{1.833pt}{0.118pt}}
\multiput(1117.00,345.92)(15.195,12.000){2}{\rule{0.917pt}{1.000pt}}
\put(464.0,163.0){\rule[-0.500pt]{13.972pt}{1.000pt}}
\end{picture}
\vspace{0.4cm}
\caption{Illustrative plot for changes of two $D^0$-$\bar{D}^0$ mixing
observables with proper time $t$, where $x^{~}_D =0.05$,
$y^{~}_D =0$, $\delta_{K\pi}=0$ and $\phi_D =\pi/2$ have been taken.}
\end{figure}

Next we consider the $D^0$-$\bar{D}^0$ mixing signals in neutral $D$ decays
to $CP$ eigenstates $K_S \pi^0$ and $K_L \pi^0$, where tiny $CP$-violating
effects induced by $D^0$-$\bar{D}^0$ mixing ($\Delta_D$) and $K^0$-$\bar{K}^0$ mixing 
($\epsilon^{~}_K$) are neglected. Since $D^0\rightarrow K^0\pi^0$ is doubly
Cabibbo-suppressed in contrast with the Cabibbo-allowed transition 
$D^0\rightarrow \bar{K}^0\pi^0$, we define a ratio
\begin{equation}
R^{\prime}_{\rm\scriptstyle DCSD} \; \equiv \; \left | \frac{\langle K^0\pi^0|{\cal H}_{\rm eff}
|D^0\rangle}{\langle \bar{K}^0\pi^0|{\cal H}_{\rm eff}|D^0\rangle} \right |^2 \; ,
\end{equation}
analogous to $R_{\rm\scriptstyle DCSD}$ defined above. Obviously $R^{\prime}_{\rm\scriptstyle
DCSD}$ and $R_{\rm\scriptstyle DCSD}$ are comparable in magnitude.
To a good degree of accuracy,  the
time-dependent decay rates of $D^0(t)\rightarrow K_{S,L} + \pi^0$ and
$\bar{D}^0(t)\rightarrow K_{S,L} + \pi^0$ are independent of $R^{\prime}_{\rm\scriptstyle DCSD}$ \cite{Xing97}:
\begin{eqnarray}
\frac{{\cal R} [\bar{D}^0 (t) \rightarrow K_S\pi^0]}{{\cal R}
[D^0 (t) \rightarrow K_S\pi^0]} & = & \frac{ 2 ~ + ~ 2 \left (y^{~}_D \cos\phi_D -
x^{~}_D \sin\phi_D \right ) (\Gamma t) ~ + ~ y^2_D (\Gamma t)^2 }
{2 ~ + ~ 2 \left (y^{~}_D \cos\phi_D + x^{~}_D \sin\phi_D \right ) (\Gamma t) ~ + ~
y^2_D (\Gamma t)^2 } \; , \nonumber \\
\frac{{\cal R} [\bar{D}^0 (t) \rightarrow K_L\pi^0]}{{\cal R}
[D^0 (t) \rightarrow K_L\pi^0]} & = & \frac{ 2 ~ - ~ 2 \left (y^{~}_D \cos\phi_D -
x^{~}_D \sin\phi_D \right ) (\Gamma t) ~ + ~ y^2_D (\Gamma t)^2 }
{2 ~ - ~ 2 \left (y^{~}_D \cos\phi_D + x^{~}_D \sin\phi_D \right ) (\Gamma t) ~ + ~
y^2_D (\Gamma t)^2 } \; ,
\end{eqnarray}
where a tiny CKM phase from the direct transition amplitude has been 
neglected. One can see that the deviation of each of the above two observables 
from unity measures nonvanishing $x^{~}_D \sin\phi_D$, while the difference
between them measures the magnitude of $y^{~}_D \cos\phi_D - x^{~}_D \sin\phi_D$.
Thus some useful information about $x^{~}_D$, $y^{~}_D$ and $\phi_D$ should
be achievable from the experimental study of those decay modes. 
We typically take (a) $\phi_D =\pi/4$, $x^{~}_D = y^{~}_D = 0.05$
and (b) $\phi_D =\pi/2$, $x^{~}_D = 0.05$, $y^{~}_D = 0$, to plot changes of
${\cal R}[\bar{D}^0(t)\rightarrow K_{S,L}+\pi^0] / {\cal R}[D^0(t)\rightarrow
K_{S,L}+ \pi^0]$ with proper time $t$ in Fig. 2 for the purpose of illustration.
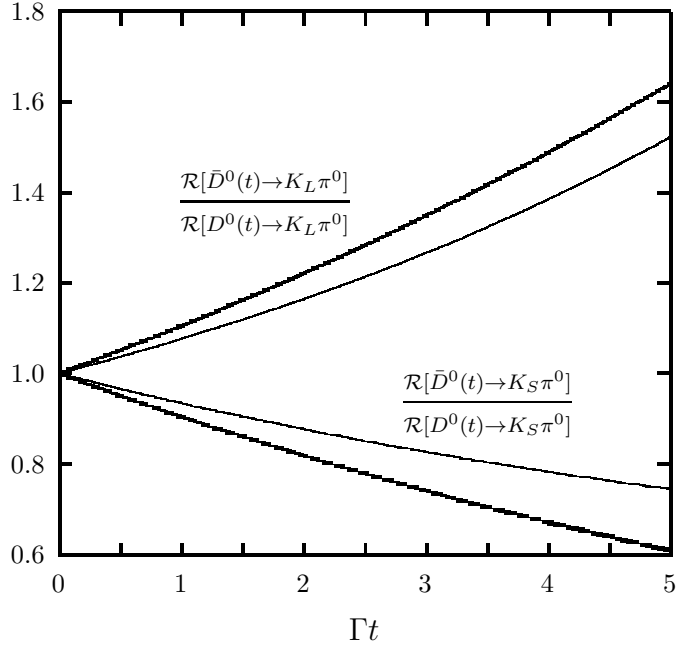
\begin{figure}
\setlength{\unitlength}{0.240900pt}
\ifx\plotpoint\undefined\newsavebox{\plotpoint}\fi
\sbox{\plotpoint}{\rule[-0.500pt]{1.000pt}{1.000pt}}%
\begin{picture}(1200,990)(-350,0)
\font\gnuplot=cmr10 at 10pt
\gnuplot
\sbox{\plotpoint}{\rule[-0.500pt]{1.000pt}{1.000pt}}%
\put(176.0,113.0){\rule[-0.500pt]{1.000pt}{205.729pt}}
\put(176.0,113.0){\rule[-0.500pt]{4.818pt}{1.000pt}}
\put(154,113){\makebox(0,0)[r]{0.6}}
\put(1116.0,113.0){\rule[-0.500pt]{4.818pt}{1.000pt}}
\put(176.0,255.0){\rule[-0.500pt]{4.818pt}{1.000pt}}
\put(154,255){\makebox(0,0)[r]{0.8}}
\put(1116.0,255.0){\rule[-0.500pt]{4.818pt}{1.000pt}}
\put(176.0,398.0){\rule[-0.500pt]{4.818pt}{1.000pt}}
\put(154,398){\makebox(0,0)[r]{1.0}}
\put(1116.0,398.0){\rule[-0.500pt]{4.818pt}{1.000pt}}
\put(176.0,540.0){\rule[-0.500pt]{4.818pt}{1.000pt}}
\put(154,540){\makebox(0,0)[r]{1.2}}
\put(1116.0,540.0){\rule[-0.500pt]{4.818pt}{1.000pt}}
\put(176.0,682.0){\rule[-0.500pt]{4.818pt}{1.000pt}}
\put(154,682){\makebox(0,0)[r]{1.4}}
\put(1116.0,682.0){\rule[-0.500pt]{4.818pt}{1.000pt}}
\put(176.0,825.0){\rule[-0.500pt]{4.818pt}{1.000pt}}
\put(154,825){\makebox(0,0)[r]{1.6}}
\put(1116.0,825.0){\rule[-0.500pt]{4.818pt}{1.000pt}}
\put(176.0,967.0){\rule[-0.500pt]{4.818pt}{1.000pt}}
\put(154,967){\makebox(0,0)[r]{1.8}}
\put(1116.0,967.0){\rule[-0.500pt]{4.818pt}{1.000pt}}
\put(176.0,113.0){\rule[-0.500pt]{1.000pt}{4.818pt}}
\put(176,68){\makebox(0,0){0}}
\put(176.0,947.0){\rule[-0.500pt]{1.000pt}{4.818pt}}
\put(272.0,113.0){\rule[-0.500pt]{1.000pt}{4.818pt}}
\put(272.0,947.0){\rule[-0.500pt]{1.000pt}{4.818pt}}
\put(368.0,113.0){\rule[-0.500pt]{1.000pt}{4.818pt}}
\put(368,68){\makebox(0,0){1}}
\put(368.0,947.0){\rule[-0.500pt]{1.000pt}{4.818pt}}
\put(464.0,113.0){\rule[-0.500pt]{1.000pt}{4.818pt}}
\put(464.0,947.0){\rule[-0.500pt]{1.000pt}{4.818pt}}
\put(560.0,113.0){\rule[-0.500pt]{1.000pt}{4.818pt}}
\put(560,68){\makebox(0,0){2}}
\put(560.0,947.0){\rule[-0.500pt]{1.000pt}{4.818pt}}
\put(656.0,113.0){\rule[-0.500pt]{1.000pt}{4.818pt}}
\put(656.0,947.0){\rule[-0.500pt]{1.000pt}{4.818pt}}
\put(752.0,113.0){\rule[-0.500pt]{1.000pt}{4.818pt}}
\put(752,68){\makebox(0,0){3}}
\put(752.0,947.0){\rule[-0.500pt]{1.000pt}{4.818pt}}
\put(848.0,113.0){\rule[-0.500pt]{1.000pt}{4.818pt}}
\put(848.0,947.0){\rule[-0.500pt]{1.000pt}{4.818pt}}
\put(944.0,113.0){\rule[-0.500pt]{1.000pt}{4.818pt}}
\put(944,68){\makebox(0,0){4}}
\put(944.0,947.0){\rule[-0.500pt]{1.000pt}{4.818pt}}
\put(1040.0,113.0){\rule[-0.500pt]{1.000pt}{4.818pt}}
\put(1040.0,947.0){\rule[-0.500pt]{1.000pt}{4.818pt}}
\put(1136.0,113.0){\rule[-0.500pt]{1.000pt}{4.818pt}}
\put(1136,68){\makebox(0,0){5}}
\put(1136.0,947.0){\rule[-0.500pt]{1.000pt}{4.818pt}}
\put(176.0,113.0){\rule[-0.500pt]{231.264pt}{1.000pt}}
\put(1136.0,113.0){\rule[-0.500pt]{1.000pt}{205.729pt}}
\put(176.0,967.0){\rule[-0.500pt]{231.264pt}{1.000pt}}
\put(656,-3){\makebox(0,0){$\Gamma t$}}
\put(500,665){\makebox(0,0){$\displaystyle\frac{\scriptstyle
{\cal R}[\bar{D}^0(t)\rightarrow K_L \pi^0]}{\scriptstyle 
{\cal R}[D^0(t)\rightarrow K_L \pi^0]}$}}
\put(850,350){\makebox(0,0){$\displaystyle\frac{\scriptstyle
{\cal R}[\bar{D}^0(t)\rightarrow K_S \pi^0]}{\scriptstyle
{\cal R}[D^0(t)\rightarrow K_S \pi^0]}$}}

\put(176.0,113.0){\rule[-0.500pt]{1.000pt}{205.729pt}}
\put(176,398){\usebox{\plotpoint}}
\multiput(176.00,395.69)(1.339,-0.475){6}{\rule{2.964pt}{0.114pt}}
\multiput(176.00,395.92)(12.847,-7.000){2}{\rule{1.482pt}{1.000pt}}
\multiput(195.00,388.69)(1.339,-0.475){6}{\rule{2.964pt}{0.114pt}}
\multiput(195.00,388.92)(12.847,-7.000){2}{\rule{1.482pt}{1.000pt}}
\multiput(214.00,381.69)(1.420,-0.475){6}{\rule{3.107pt}{0.114pt}}
\multiput(214.00,381.92)(13.551,-7.000){2}{\rule{1.554pt}{1.000pt}}
\multiput(234.00,374.69)(1.339,-0.475){6}{\rule{2.964pt}{0.114pt}}
\multiput(234.00,374.92)(12.847,-7.000){2}{\rule{1.482pt}{1.000pt}}
\multiput(253.00,367.69)(1.339,-0.475){6}{\rule{2.964pt}{0.114pt}}
\multiput(253.00,367.92)(12.847,-7.000){2}{\rule{1.482pt}{1.000pt}}
\multiput(272.00,360.69)(1.339,-0.475){6}{\rule{2.964pt}{0.114pt}}
\multiput(272.00,360.92)(12.847,-7.000){2}{\rule{1.482pt}{1.000pt}}
\multiput(291.00,353.69)(1.606,-0.462){4}{\rule{3.417pt}{0.111pt}}
\multiput(291.00,353.92)(11.909,-6.000){2}{\rule{1.708pt}{1.000pt}}
\multiput(310.00,347.69)(1.420,-0.475){6}{\rule{3.107pt}{0.114pt}}
\multiput(310.00,347.92)(13.551,-7.000){2}{\rule{1.554pt}{1.000pt}}
\multiput(330.00,340.69)(1.339,-0.475){6}{\rule{2.964pt}{0.114pt}}
\multiput(330.00,340.92)(12.847,-7.000){2}{\rule{1.482pt}{1.000pt}}
\multiput(349.00,333.69)(1.606,-0.462){4}{\rule{3.417pt}{0.111pt}}
\multiput(349.00,333.92)(11.909,-6.000){2}{\rule{1.708pt}{1.000pt}}
\multiput(368.00,327.69)(1.606,-0.462){4}{\rule{3.417pt}{0.111pt}}
\multiput(368.00,327.92)(11.909,-6.000){2}{\rule{1.708pt}{1.000pt}}
\multiput(387.00,321.69)(1.339,-0.475){6}{\rule{2.964pt}{0.114pt}}
\multiput(387.00,321.92)(12.847,-7.000){2}{\rule{1.482pt}{1.000pt}}
\multiput(406.00,314.69)(1.708,-0.462){4}{\rule{3.583pt}{0.111pt}}
\multiput(406.00,314.92)(12.563,-6.000){2}{\rule{1.792pt}{1.000pt}}
\multiput(426.00,308.69)(1.606,-0.462){4}{\rule{3.417pt}{0.111pt}}
\multiput(426.00,308.92)(11.909,-6.000){2}{\rule{1.708pt}{1.000pt}}
\multiput(445.00,302.69)(1.606,-0.462){4}{\rule{3.417pt}{0.111pt}}
\multiput(445.00,302.92)(11.909,-6.000){2}{\rule{1.708pt}{1.000pt}}
\multiput(464.00,296.69)(1.606,-0.462){4}{\rule{3.417pt}{0.111pt}}
\multiput(464.00,296.92)(11.909,-6.000){2}{\rule{1.708pt}{1.000pt}}
\multiput(483.00,290.69)(1.606,-0.462){4}{\rule{3.417pt}{0.111pt}}
\multiput(483.00,290.92)(11.909,-6.000){2}{\rule{1.708pt}{1.000pt}}
\multiput(502.00,284.69)(1.708,-0.462){4}{\rule{3.583pt}{0.111pt}}
\multiput(502.00,284.92)(12.563,-6.000){2}{\rule{1.792pt}{1.000pt}}
\multiput(522.00,278.69)(1.606,-0.462){4}{\rule{3.417pt}{0.111pt}}
\multiput(522.00,278.92)(11.909,-6.000){2}{\rule{1.708pt}{1.000pt}}
\multiput(541.00,272.69)(1.606,-0.462){4}{\rule{3.417pt}{0.111pt}}
\multiput(541.00,272.92)(11.909,-6.000){2}{\rule{1.708pt}{1.000pt}}
\multiput(560.00,266.69)(1.606,-0.462){4}{\rule{3.417pt}{0.111pt}}
\multiput(560.00,266.92)(11.909,-6.000){2}{\rule{1.708pt}{1.000pt}}
\multiput(579.00,260.69)(1.606,-0.462){4}{\rule{3.417pt}{0.111pt}}
\multiput(579.00,260.92)(11.909,-6.000){2}{\rule{1.708pt}{1.000pt}}
\multiput(598.00,254.71)(2.358,-0.424){2}{\rule{4.250pt}{0.102pt}}
\multiput(598.00,254.92)(11.179,-5.000){2}{\rule{2.125pt}{1.000pt}}
\multiput(618.00,249.69)(1.606,-0.462){4}{\rule{3.417pt}{0.111pt}}
\multiput(618.00,249.92)(11.909,-6.000){2}{\rule{1.708pt}{1.000pt}}
\multiput(637.00,243.71)(2.188,-0.424){2}{\rule{4.050pt}{0.102pt}}
\multiput(637.00,243.92)(10.594,-5.000){2}{\rule{2.025pt}{1.000pt}}
\multiput(656.00,238.69)(1.606,-0.462){4}{\rule{3.417pt}{0.111pt}}
\multiput(656.00,238.92)(11.909,-6.000){2}{\rule{1.708pt}{1.000pt}}
\multiput(675.00,232.71)(2.188,-0.424){2}{\rule{4.050pt}{0.102pt}}
\multiput(675.00,232.92)(10.594,-5.000){2}{\rule{2.025pt}{1.000pt}}
\multiput(694.00,227.69)(1.708,-0.462){4}{\rule{3.583pt}{0.111pt}}
\multiput(694.00,227.92)(12.563,-6.000){2}{\rule{1.792pt}{1.000pt}}
\multiput(714.00,221.71)(2.188,-0.424){2}{\rule{4.050pt}{0.102pt}}
\multiput(714.00,221.92)(10.594,-5.000){2}{\rule{2.025pt}{1.000pt}}
\multiput(733.00,216.71)(2.188,-0.424){2}{\rule{4.050pt}{0.102pt}}
\multiput(733.00,216.92)(10.594,-5.000){2}{\rule{2.025pt}{1.000pt}}
\multiput(752.00,211.71)(2.188,-0.424){2}{\rule{4.050pt}{0.102pt}}
\multiput(752.00,211.92)(10.594,-5.000){2}{\rule{2.025pt}{1.000pt}}
\multiput(771.00,206.69)(1.606,-0.462){4}{\rule{3.417pt}{0.111pt}}
\multiput(771.00,206.92)(11.909,-6.000){2}{\rule{1.708pt}{1.000pt}}
\multiput(790.00,200.71)(2.358,-0.424){2}{\rule{4.250pt}{0.102pt}}
\multiput(790.00,200.92)(11.179,-5.000){2}{\rule{2.125pt}{1.000pt}}
\multiput(810.00,195.71)(2.188,-0.424){2}{\rule{4.050pt}{0.102pt}}
\multiput(810.00,195.92)(10.594,-5.000){2}{\rule{2.025pt}{1.000pt}}
\multiput(829.00,190.71)(2.188,-0.424){2}{\rule{4.050pt}{0.102pt}}
\multiput(829.00,190.92)(10.594,-5.000){2}{\rule{2.025pt}{1.000pt}}
\multiput(848.00,185.71)(2.188,-0.424){2}{\rule{4.050pt}{0.102pt}}
\multiput(848.00,185.92)(10.594,-5.000){2}{\rule{2.025pt}{1.000pt}}
\put(867,178.92){\rule{4.577pt}{1.000pt}}
\multiput(867.00,180.92)(9.500,-4.000){2}{\rule{2.289pt}{1.000pt}}
\multiput(886.00,176.71)(2.358,-0.424){2}{\rule{4.250pt}{0.102pt}}
\multiput(886.00,176.92)(11.179,-5.000){2}{\rule{2.125pt}{1.000pt}}
\multiput(906.00,171.71)(2.188,-0.424){2}{\rule{4.050pt}{0.102pt}}
\multiput(906.00,171.92)(10.594,-5.000){2}{\rule{2.025pt}{1.000pt}}
\multiput(925.00,166.71)(2.188,-0.424){2}{\rule{4.050pt}{0.102pt}}
\multiput(925.00,166.92)(10.594,-5.000){2}{\rule{2.025pt}{1.000pt}}
\put(944,159.92){\rule{4.577pt}{1.000pt}}
\multiput(944.00,161.92)(9.500,-4.000){2}{\rule{2.289pt}{1.000pt}}
\multiput(963.00,157.71)(2.188,-0.424){2}{\rule{4.050pt}{0.102pt}}
\multiput(963.00,157.92)(10.594,-5.000){2}{\rule{2.025pt}{1.000pt}}
\put(982,150.92){\rule{4.818pt}{1.000pt}}
\multiput(982.00,152.92)(10.000,-4.000){2}{\rule{2.409pt}{1.000pt}}
\multiput(1002.00,148.71)(2.188,-0.424){2}{\rule{4.050pt}{0.102pt}}
\multiput(1002.00,148.92)(10.594,-5.000){2}{\rule{2.025pt}{1.000pt}}
\put(1021,141.92){\rule{4.577pt}{1.000pt}}
\multiput(1021.00,143.92)(9.500,-4.000){2}{\rule{2.289pt}{1.000pt}}
\multiput(1040.00,139.71)(2.188,-0.424){2}{\rule{4.050pt}{0.102pt}}
\multiput(1040.00,139.92)(10.594,-5.000){2}{\rule{2.025pt}{1.000pt}}
\put(1059,132.92){\rule{4.577pt}{1.000pt}}
\multiput(1059.00,134.92)(9.500,-4.000){2}{\rule{2.289pt}{1.000pt}}
\multiput(1078.00,130.71)(2.358,-0.424){2}{\rule{4.250pt}{0.102pt}}
\multiput(1078.00,130.92)(11.179,-5.000){2}{\rule{2.125pt}{1.000pt}}
\put(1098,123.92){\rule{4.577pt}{1.000pt}}
\multiput(1098.00,125.92)(9.500,-4.000){2}{\rule{2.289pt}{1.000pt}}
\put(1117,119.92){\rule{4.577pt}{1.000pt}}
\multiput(1117.00,121.92)(9.500,-4.000){2}{\rule{2.289pt}{1.000pt}}
\put(176,398){\usebox{\plotpoint}}
\multiput(176.00,399.84)(1.339,0.475){6}{\rule{2.964pt}{0.114pt}}
\multiput(176.00,395.92)(12.847,7.000){2}{\rule{1.482pt}{1.000pt}}
\multiput(195.00,406.84)(1.339,0.475){6}{\rule{2.964pt}{0.114pt}}
\multiput(195.00,402.92)(12.847,7.000){2}{\rule{1.482pt}{1.000pt}}
\multiput(214.00,413.84)(1.420,0.475){6}{\rule{3.107pt}{0.114pt}}
\multiput(214.00,409.92)(13.551,7.000){2}{\rule{1.554pt}{1.000pt}}
\multiput(234.00,420.83)(1.158,0.481){8}{\rule{2.625pt}{0.116pt}}
\multiput(234.00,416.92)(13.552,8.000){2}{\rule{1.313pt}{1.000pt}}
\multiput(253.00,428.84)(1.339,0.475){6}{\rule{2.964pt}{0.114pt}}
\multiput(253.00,424.92)(12.847,7.000){2}{\rule{1.482pt}{1.000pt}}
\multiput(272.00,435.83)(1.158,0.481){8}{\rule{2.625pt}{0.116pt}}
\multiput(272.00,431.92)(13.552,8.000){2}{\rule{1.313pt}{1.000pt}}
\multiput(291.00,443.84)(1.339,0.475){6}{\rule{2.964pt}{0.114pt}}
\multiput(291.00,439.92)(12.847,7.000){2}{\rule{1.482pt}{1.000pt}}
\multiput(310.00,450.83)(1.226,0.481){8}{\rule{2.750pt}{0.116pt}}
\multiput(310.00,446.92)(14.292,8.000){2}{\rule{1.375pt}{1.000pt}}
\multiput(330.00,458.83)(1.158,0.481){8}{\rule{2.625pt}{0.116pt}}
\multiput(330.00,454.92)(13.552,8.000){2}{\rule{1.313pt}{1.000pt}}
\multiput(349.00,466.84)(1.339,0.475){6}{\rule{2.964pt}{0.114pt}}
\multiput(349.00,462.92)(12.847,7.000){2}{\rule{1.482pt}{1.000pt}}
\multiput(368.00,473.83)(1.158,0.481){8}{\rule{2.625pt}{0.116pt}}
\multiput(368.00,469.92)(13.552,8.000){2}{\rule{1.313pt}{1.000pt}}
\multiput(387.00,481.83)(1.158,0.481){8}{\rule{2.625pt}{0.116pt}}
\multiput(387.00,477.92)(13.552,8.000){2}{\rule{1.313pt}{1.000pt}}
\multiput(406.00,489.83)(1.226,0.481){8}{\rule{2.750pt}{0.116pt}}
\multiput(406.00,485.92)(14.292,8.000){2}{\rule{1.375pt}{1.000pt}}
\multiput(426.00,497.83)(1.022,0.485){10}{\rule{2.361pt}{0.117pt}}
\multiput(426.00,493.92)(14.099,9.000){2}{\rule{1.181pt}{1.000pt}}
\multiput(445.00,506.83)(1.158,0.481){8}{\rule{2.625pt}{0.116pt}}
\multiput(445.00,502.92)(13.552,8.000){2}{\rule{1.313pt}{1.000pt}}
\multiput(464.00,514.83)(1.158,0.481){8}{\rule{2.625pt}{0.116pt}}
\multiput(464.00,510.92)(13.552,8.000){2}{\rule{1.313pt}{1.000pt}}
\multiput(483.00,522.83)(1.158,0.481){8}{\rule{2.625pt}{0.116pt}}
\multiput(483.00,518.92)(13.552,8.000){2}{\rule{1.313pt}{1.000pt}}
\multiput(502.00,530.83)(1.082,0.485){10}{\rule{2.472pt}{0.117pt}}
\multiput(502.00,526.92)(14.869,9.000){2}{\rule{1.236pt}{1.000pt}}
\multiput(522.00,539.83)(1.158,0.481){8}{\rule{2.625pt}{0.116pt}}
\multiput(522.00,535.92)(13.552,8.000){2}{\rule{1.313pt}{1.000pt}}
\multiput(541.00,547.83)(1.022,0.485){10}{\rule{2.361pt}{0.117pt}}
\multiput(541.00,543.92)(14.099,9.000){2}{\rule{1.181pt}{1.000pt}}
\multiput(560.00,556.83)(1.022,0.485){10}{\rule{2.361pt}{0.117pt}}
\multiput(560.00,552.92)(14.099,9.000){2}{\rule{1.181pt}{1.000pt}}
\multiput(579.00,565.83)(1.158,0.481){8}{\rule{2.625pt}{0.116pt}}
\multiput(579.00,561.92)(13.552,8.000){2}{\rule{1.313pt}{1.000pt}}
\multiput(598.00,573.83)(1.082,0.485){10}{\rule{2.472pt}{0.117pt}}
\multiput(598.00,569.92)(14.869,9.000){2}{\rule{1.236pt}{1.000pt}}
\multiput(618.00,582.83)(1.022,0.485){10}{\rule{2.361pt}{0.117pt}}
\multiput(618.00,578.92)(14.099,9.000){2}{\rule{1.181pt}{1.000pt}}
\multiput(637.00,591.83)(1.022,0.485){10}{\rule{2.361pt}{0.117pt}}
\multiput(637.00,587.92)(14.099,9.000){2}{\rule{1.181pt}{1.000pt}}
\multiput(656.00,600.83)(1.022,0.485){10}{\rule{2.361pt}{0.117pt}}
\multiput(656.00,596.92)(14.099,9.000){2}{\rule{1.181pt}{1.000pt}}
\multiput(675.00,609.83)(1.022,0.485){10}{\rule{2.361pt}{0.117pt}}
\multiput(675.00,605.92)(14.099,9.000){2}{\rule{1.181pt}{1.000pt}}
\multiput(694.00,618.83)(0.969,0.487){12}{\rule{2.250pt}{0.117pt}}
\multiput(694.00,614.92)(15.330,10.000){2}{\rule{1.125pt}{1.000pt}}
\multiput(714.00,628.83)(1.022,0.485){10}{\rule{2.361pt}{0.117pt}}
\multiput(714.00,624.92)(14.099,9.000){2}{\rule{1.181pt}{1.000pt}}
\multiput(733.00,637.83)(0.916,0.487){12}{\rule{2.150pt}{0.117pt}}
\multiput(733.00,633.92)(14.538,10.000){2}{\rule{1.075pt}{1.000pt}}
\multiput(752.00,647.83)(1.022,0.485){10}{\rule{2.361pt}{0.117pt}}
\multiput(752.00,643.92)(14.099,9.000){2}{\rule{1.181pt}{1.000pt}}
\multiput(771.00,656.83)(0.916,0.487){12}{\rule{2.150pt}{0.117pt}}
\multiput(771.00,652.92)(14.538,10.000){2}{\rule{1.075pt}{1.000pt}}
\multiput(790.00,666.83)(1.082,0.485){10}{\rule{2.472pt}{0.117pt}}
\multiput(790.00,662.92)(14.869,9.000){2}{\rule{1.236pt}{1.000pt}}
\multiput(810.00,675.83)(0.916,0.487){12}{\rule{2.150pt}{0.117pt}}
\multiput(810.00,671.92)(14.538,10.000){2}{\rule{1.075pt}{1.000pt}}
\multiput(829.00,685.83)(0.916,0.487){12}{\rule{2.150pt}{0.117pt}}
\multiput(829.00,681.92)(14.538,10.000){2}{\rule{1.075pt}{1.000pt}}
\multiput(848.00,695.83)(0.916,0.487){12}{\rule{2.150pt}{0.117pt}}
\multiput(848.00,691.92)(14.538,10.000){2}{\rule{1.075pt}{1.000pt}}
\multiput(867.00,705.83)(0.916,0.487){12}{\rule{2.150pt}{0.117pt}}
\multiput(867.00,701.92)(14.538,10.000){2}{\rule{1.075pt}{1.000pt}}
\multiput(886.00,715.83)(0.969,0.487){12}{\rule{2.250pt}{0.117pt}}
\multiput(886.00,711.92)(15.330,10.000){2}{\rule{1.125pt}{1.000pt}}
\multiput(906.00,725.83)(0.916,0.487){12}{\rule{2.150pt}{0.117pt}}
\multiput(906.00,721.92)(14.538,10.000){2}{\rule{1.075pt}{1.000pt}}
\multiput(925.00,735.83)(0.830,0.489){14}{\rule{1.977pt}{0.118pt}}
\multiput(925.00,731.92)(14.896,11.000){2}{\rule{0.989pt}{1.000pt}}
\multiput(944.00,746.83)(0.916,0.487){12}{\rule{2.150pt}{0.117pt}}
\multiput(944.00,742.92)(14.538,10.000){2}{\rule{1.075pt}{1.000pt}}
\multiput(963.00,756.83)(0.830,0.489){14}{\rule{1.977pt}{0.118pt}}
\multiput(963.00,752.92)(14.896,11.000){2}{\rule{0.989pt}{1.000pt}}
\multiput(982.00,767.83)(0.969,0.487){12}{\rule{2.250pt}{0.117pt}}
\multiput(982.00,763.92)(15.330,10.000){2}{\rule{1.125pt}{1.000pt}}
\multiput(1002.00,777.83)(0.830,0.489){14}{\rule{1.977pt}{0.118pt}}
\multiput(1002.00,773.92)(14.896,11.000){2}{\rule{0.989pt}{1.000pt}}
\multiput(1021.00,788.83)(0.830,0.489){14}{\rule{1.977pt}{0.118pt}}
\multiput(1021.00,784.92)(14.896,11.000){2}{\rule{0.989pt}{1.000pt}}
\multiput(1040.00,799.83)(0.830,0.489){14}{\rule{1.977pt}{0.118pt}}
\multiput(1040.00,795.92)(14.896,11.000){2}{\rule{0.989pt}{1.000pt}}
\multiput(1059.00,810.83)(0.830,0.489){14}{\rule{1.977pt}{0.118pt}}
\multiput(1059.00,806.92)(14.896,11.000){2}{\rule{0.989pt}{1.000pt}}
\multiput(1078.00,821.83)(0.878,0.489){14}{\rule{2.068pt}{0.118pt}}
\multiput(1078.00,817.92)(15.707,11.000){2}{\rule{1.034pt}{1.000pt}}
\multiput(1098.00,832.83)(0.830,0.489){14}{\rule{1.977pt}{0.118pt}}
\multiput(1098.00,828.92)(14.896,11.000){2}{\rule{0.989pt}{1.000pt}}
\multiput(1117.00,843.83)(0.830,0.489){14}{\rule{1.977pt}{0.118pt}}
\multiput(1117.00,839.92)(14.896,11.000){2}{\rule{0.989pt}{1.000pt}}
\sbox{\plotpoint}{\rule[-0.175pt]{0.350pt}{0.350pt}}%
\put(176,398){\usebox{\plotpoint}}
\multiput(176.00,397.02)(2.186,-0.507){7}{\rule{1.418pt}{0.122pt}}
\multiput(176.00,397.27)(16.058,-5.000){2}{\rule{0.709pt}{0.350pt}}
\multiput(195.00,392.02)(2.186,-0.507){7}{\rule{1.418pt}{0.122pt}}
\multiput(195.00,392.27)(16.058,-5.000){2}{\rule{0.709pt}{0.350pt}}
\multiput(214.00,387.02)(2.304,-0.507){7}{\rule{1.487pt}{0.122pt}}
\multiput(214.00,387.27)(16.913,-5.000){2}{\rule{0.744pt}{0.350pt}}
\multiput(234.00,382.02)(2.186,-0.507){7}{\rule{1.418pt}{0.122pt}}
\multiput(234.00,382.27)(16.058,-5.000){2}{\rule{0.709pt}{0.350pt}}
\multiput(253.00,377.02)(2.186,-0.507){7}{\rule{1.418pt}{0.122pt}}
\multiput(253.00,377.27)(16.058,-5.000){2}{\rule{0.709pt}{0.350pt}}
\multiput(272.00,372.02)(2.934,-0.509){5}{\rule{1.750pt}{0.123pt}}
\multiput(272.00,372.27)(15.368,-4.000){2}{\rule{0.875pt}{0.350pt}}
\multiput(291.00,368.02)(2.186,-0.507){7}{\rule{1.418pt}{0.122pt}}
\multiput(291.00,368.27)(16.058,-5.000){2}{\rule{0.709pt}{0.350pt}}
\multiput(310.00,363.02)(3.093,-0.509){5}{\rule{1.837pt}{0.123pt}}
\multiput(310.00,363.27)(16.186,-4.000){2}{\rule{0.919pt}{0.350pt}}
\multiput(330.00,359.02)(2.186,-0.507){7}{\rule{1.418pt}{0.122pt}}
\multiput(330.00,359.27)(16.058,-5.000){2}{\rule{0.709pt}{0.350pt}}
\multiput(349.00,354.02)(2.934,-0.509){5}{\rule{1.750pt}{0.123pt}}
\multiput(349.00,354.27)(15.368,-4.000){2}{\rule{0.875pt}{0.350pt}}
\multiput(368.00,350.02)(2.186,-0.507){7}{\rule{1.418pt}{0.122pt}}
\multiput(368.00,350.27)(16.058,-5.000){2}{\rule{0.709pt}{0.350pt}}
\multiput(387.00,345.02)(2.934,-0.509){5}{\rule{1.750pt}{0.123pt}}
\multiput(387.00,345.27)(15.368,-4.000){2}{\rule{0.875pt}{0.350pt}}
\multiput(406.00,341.02)(3.093,-0.509){5}{\rule{1.837pt}{0.123pt}}
\multiput(406.00,341.27)(16.186,-4.000){2}{\rule{0.919pt}{0.350pt}}
\multiput(426.00,337.02)(2.934,-0.509){5}{\rule{1.750pt}{0.123pt}}
\multiput(426.00,337.27)(15.368,-4.000){2}{\rule{0.875pt}{0.350pt}}
\multiput(445.00,333.02)(2.934,-0.509){5}{\rule{1.750pt}{0.123pt}}
\multiput(445.00,333.27)(15.368,-4.000){2}{\rule{0.875pt}{0.350pt}}
\multiput(464.00,329.02)(2.934,-0.509){5}{\rule{1.750pt}{0.123pt}}
\multiput(464.00,329.27)(15.368,-4.000){2}{\rule{0.875pt}{0.350pt}}
\multiput(483.00,325.02)(2.934,-0.509){5}{\rule{1.750pt}{0.123pt}}
\multiput(483.00,325.27)(15.368,-4.000){2}{\rule{0.875pt}{0.350pt}}
\multiput(502.00,321.02)(3.093,-0.509){5}{\rule{1.837pt}{0.123pt}}
\multiput(502.00,321.27)(16.186,-4.000){2}{\rule{0.919pt}{0.350pt}}
\multiput(522.00,317.02)(2.934,-0.509){5}{\rule{1.750pt}{0.123pt}}
\multiput(522.00,317.27)(15.368,-4.000){2}{\rule{0.875pt}{0.350pt}}
\multiput(541.00,313.02)(2.934,-0.509){5}{\rule{1.750pt}{0.123pt}}
\multiput(541.00,313.27)(15.368,-4.000){2}{\rule{0.875pt}{0.350pt}}
\multiput(560.00,309.02)(2.934,-0.509){5}{\rule{1.750pt}{0.123pt}}
\multiput(560.00,309.27)(15.368,-4.000){2}{\rule{0.875pt}{0.350pt}}
\multiput(579.00,305.02)(2.934,-0.509){5}{\rule{1.750pt}{0.123pt}}
\multiput(579.00,305.27)(15.368,-4.000){2}{\rule{0.875pt}{0.350pt}}
\multiput(598.00,301.02)(4.975,-0.516){3}{\rule{2.421pt}{0.124pt}}
\multiput(598.00,301.27)(14.975,-3.000){2}{\rule{1.210pt}{0.350pt}}
\multiput(618.00,298.02)(2.934,-0.509){5}{\rule{1.750pt}{0.123pt}}
\multiput(618.00,298.27)(15.368,-4.000){2}{\rule{0.875pt}{0.350pt}}
\multiput(637.00,294.02)(2.934,-0.509){5}{\rule{1.750pt}{0.123pt}}
\multiput(637.00,294.27)(15.368,-4.000){2}{\rule{0.875pt}{0.350pt}}
\multiput(656.00,290.02)(4.718,-0.516){3}{\rule{2.304pt}{0.124pt}}
\multiput(656.00,290.27)(14.218,-3.000){2}{\rule{1.152pt}{0.350pt}}
\multiput(675.00,287.02)(2.934,-0.509){5}{\rule{1.750pt}{0.123pt}}
\multiput(675.00,287.27)(15.368,-4.000){2}{\rule{0.875pt}{0.350pt}}
\multiput(694.00,283.02)(4.975,-0.516){3}{\rule{2.421pt}{0.124pt}}
\multiput(694.00,283.27)(14.975,-3.000){2}{\rule{1.210pt}{0.350pt}}
\multiput(714.00,280.02)(4.718,-0.516){3}{\rule{2.304pt}{0.124pt}}
\multiput(714.00,280.27)(14.218,-3.000){2}{\rule{1.152pt}{0.350pt}}
\multiput(733.00,277.02)(2.934,-0.509){5}{\rule{1.750pt}{0.123pt}}
\multiput(733.00,277.27)(15.368,-4.000){2}{\rule{0.875pt}{0.350pt}}
\multiput(752.00,273.02)(4.718,-0.516){3}{\rule{2.304pt}{0.124pt}}
\multiput(752.00,273.27)(14.218,-3.000){2}{\rule{1.152pt}{0.350pt}}
\multiput(771.00,270.02)(4.718,-0.516){3}{\rule{2.304pt}{0.124pt}}
\multiput(771.00,270.27)(14.218,-3.000){2}{\rule{1.152pt}{0.350pt}}
\multiput(790.00,267.02)(3.093,-0.509){5}{\rule{1.837pt}{0.123pt}}
\multiput(790.00,267.27)(16.186,-4.000){2}{\rule{0.919pt}{0.350pt}}
\multiput(810.00,263.02)(4.718,-0.516){3}{\rule{2.304pt}{0.124pt}}
\multiput(810.00,263.27)(14.218,-3.000){2}{\rule{1.152pt}{0.350pt}}
\multiput(829.00,260.02)(4.718,-0.516){3}{\rule{2.304pt}{0.124pt}}
\multiput(829.00,260.27)(14.218,-3.000){2}{\rule{1.152pt}{0.350pt}}
\multiput(848.00,257.02)(4.718,-0.516){3}{\rule{2.304pt}{0.124pt}}
\multiput(848.00,257.27)(14.218,-3.000){2}{\rule{1.152pt}{0.350pt}}
\multiput(867.00,254.02)(4.718,-0.516){3}{\rule{2.304pt}{0.124pt}}
\multiput(867.00,254.27)(14.218,-3.000){2}{\rule{1.152pt}{0.350pt}}
\multiput(886.00,251.02)(4.975,-0.516){3}{\rule{2.421pt}{0.124pt}}
\multiput(886.00,251.27)(14.975,-3.000){2}{\rule{1.210pt}{0.350pt}}
\multiput(906.00,248.02)(4.718,-0.516){3}{\rule{2.304pt}{0.124pt}}
\multiput(906.00,248.27)(14.218,-3.000){2}{\rule{1.152pt}{0.350pt}}
\multiput(925.00,245.02)(4.718,-0.516){3}{\rule{2.304pt}{0.124pt}}
\multiput(925.00,245.27)(14.218,-3.000){2}{\rule{1.152pt}{0.350pt}}
\multiput(944.00,242.02)(4.718,-0.516){3}{\rule{2.304pt}{0.124pt}}
\multiput(944.00,242.27)(14.218,-3.000){2}{\rule{1.152pt}{0.350pt}}
\multiput(963.00,239.02)(4.718,-0.516){3}{\rule{2.304pt}{0.124pt}}
\multiput(963.00,239.27)(14.218,-3.000){2}{\rule{1.152pt}{0.350pt}}
\put(982,235.27){\rule{3.588pt}{0.350pt}}
\multiput(982.00,236.27)(12.554,-2.000){2}{\rule{1.794pt}{0.350pt}}
\multiput(1002.00,234.02)(4.718,-0.516){3}{\rule{2.304pt}{0.124pt}}
\multiput(1002.00,234.27)(14.218,-3.000){2}{\rule{1.152pt}{0.350pt}}
\multiput(1021.00,231.02)(4.718,-0.516){3}{\rule{2.304pt}{0.124pt}}
\multiput(1021.00,231.27)(14.218,-3.000){2}{\rule{1.152pt}{0.350pt}}
\multiput(1040.00,228.02)(4.718,-0.516){3}{\rule{2.304pt}{0.124pt}}
\multiput(1040.00,228.27)(14.218,-3.000){2}{\rule{1.152pt}{0.350pt}}
\put(1059,224.27){\rule{3.413pt}{0.350pt}}
\multiput(1059.00,225.27)(11.917,-2.000){2}{\rule{1.706pt}{0.350pt}}
\multiput(1078.00,223.02)(4.975,-0.516){3}{\rule{2.421pt}{0.124pt}}
\multiput(1078.00,223.27)(14.975,-3.000){2}{\rule{1.210pt}{0.350pt}}
\put(1098,219.27){\rule{3.413pt}{0.350pt}}
\multiput(1098.00,220.27)(11.917,-2.000){2}{\rule{1.706pt}{0.350pt}}
\multiput(1117.00,218.02)(4.718,-0.516){3}{\rule{2.304pt}{0.124pt}}
\multiput(1117.00,218.27)(14.218,-3.000){2}{\rule{1.152pt}{0.350pt}}
\put(176,398){\usebox{\plotpoint}}
\multiput(176.00,398.47)(2.186,0.507){7}{\rule{1.418pt}{0.122pt}}
\multiput(176.00,397.27)(16.058,5.000){2}{\rule{0.709pt}{0.350pt}}
\multiput(195.00,403.47)(2.186,0.507){7}{\rule{1.418pt}{0.122pt}}
\multiput(195.00,402.27)(16.058,5.000){2}{\rule{0.709pt}{0.350pt}}
\multiput(214.00,408.47)(2.304,0.507){7}{\rule{1.487pt}{0.122pt}}
\multiput(214.00,407.27)(16.913,5.000){2}{\rule{0.744pt}{0.350pt}}
\multiput(234.00,413.47)(2.186,0.507){7}{\rule{1.418pt}{0.122pt}}
\multiput(234.00,412.27)(16.058,5.000){2}{\rule{0.709pt}{0.350pt}}
\multiput(253.00,418.47)(1.754,0.505){9}{\rule{1.196pt}{0.122pt}}
\multiput(253.00,417.27)(16.518,6.000){2}{\rule{0.598pt}{0.350pt}}
\multiput(272.00,424.47)(2.186,0.507){7}{\rule{1.418pt}{0.122pt}}
\multiput(272.00,423.27)(16.058,5.000){2}{\rule{0.709pt}{0.350pt}}
\multiput(291.00,429.47)(1.754,0.505){9}{\rule{1.196pt}{0.122pt}}
\multiput(291.00,428.27)(16.518,6.000){2}{\rule{0.598pt}{0.350pt}}
\multiput(310.00,435.47)(2.304,0.507){7}{\rule{1.487pt}{0.122pt}}
\multiput(310.00,434.27)(16.913,5.000){2}{\rule{0.744pt}{0.350pt}}
\multiput(330.00,440.47)(1.754,0.505){9}{\rule{1.196pt}{0.122pt}}
\multiput(330.00,439.27)(16.518,6.000){2}{\rule{0.598pt}{0.350pt}}
\multiput(349.00,446.47)(1.754,0.505){9}{\rule{1.196pt}{0.122pt}}
\multiput(349.00,445.27)(16.518,6.000){2}{\rule{0.598pt}{0.350pt}}
\multiput(368.00,452.47)(1.754,0.505){9}{\rule{1.196pt}{0.122pt}}
\multiput(368.00,451.27)(16.518,6.000){2}{\rule{0.598pt}{0.350pt}}
\multiput(387.00,458.47)(1.754,0.505){9}{\rule{1.196pt}{0.122pt}}
\multiput(387.00,457.27)(16.518,6.000){2}{\rule{0.598pt}{0.350pt}}
\multiput(406.00,464.47)(1.849,0.505){9}{\rule{1.254pt}{0.122pt}}
\multiput(406.00,463.27)(17.397,6.000){2}{\rule{0.627pt}{0.350pt}}
\multiput(426.00,470.47)(1.754,0.505){9}{\rule{1.196pt}{0.122pt}}
\multiput(426.00,469.27)(16.518,6.000){2}{\rule{0.598pt}{0.350pt}}
\multiput(445.00,476.47)(1.754,0.505){9}{\rule{1.196pt}{0.122pt}}
\multiput(445.00,475.27)(16.518,6.000){2}{\rule{0.598pt}{0.350pt}}
\multiput(464.00,482.47)(1.754,0.505){9}{\rule{1.196pt}{0.122pt}}
\multiput(464.00,481.27)(16.518,6.000){2}{\rule{0.598pt}{0.350pt}}
\multiput(483.00,488.47)(1.469,0.504){11}{\rule{1.038pt}{0.121pt}}
\multiput(483.00,487.27)(16.847,7.000){2}{\rule{0.519pt}{0.350pt}}
\multiput(502.00,495.47)(1.849,0.505){9}{\rule{1.254pt}{0.122pt}}
\multiput(502.00,494.27)(17.397,6.000){2}{\rule{0.627pt}{0.350pt}}
\multiput(522.00,501.47)(1.469,0.504){11}{\rule{1.038pt}{0.121pt}}
\multiput(522.00,500.27)(16.847,7.000){2}{\rule{0.519pt}{0.350pt}}
\multiput(541.00,508.47)(1.754,0.505){9}{\rule{1.196pt}{0.122pt}}
\multiput(541.00,507.27)(16.518,6.000){2}{\rule{0.598pt}{0.350pt}}
\multiput(560.00,514.47)(1.469,0.504){11}{\rule{1.038pt}{0.121pt}}
\multiput(560.00,513.27)(16.847,7.000){2}{\rule{0.519pt}{0.350pt}}
\multiput(579.00,521.47)(1.469,0.504){11}{\rule{1.038pt}{0.121pt}}
\multiput(579.00,520.27)(16.847,7.000){2}{\rule{0.519pt}{0.350pt}}
\multiput(598.00,528.47)(1.549,0.504){11}{\rule{1.087pt}{0.121pt}}
\multiput(598.00,527.27)(17.743,7.000){2}{\rule{0.544pt}{0.350pt}}
\multiput(618.00,535.47)(1.469,0.504){11}{\rule{1.038pt}{0.121pt}}
\multiput(618.00,534.27)(16.847,7.000){2}{\rule{0.519pt}{0.350pt}}
\multiput(637.00,542.47)(1.469,0.504){11}{\rule{1.038pt}{0.121pt}}
\multiput(637.00,541.27)(16.847,7.000){2}{\rule{0.519pt}{0.350pt}}
\multiput(656.00,549.47)(1.469,0.504){11}{\rule{1.038pt}{0.121pt}}
\multiput(656.00,548.27)(16.847,7.000){2}{\rule{0.519pt}{0.350pt}}
\multiput(675.00,556.47)(1.266,0.504){13}{\rule{0.919pt}{0.121pt}}
\multiput(675.00,555.27)(17.093,8.000){2}{\rule{0.459pt}{0.350pt}}
\multiput(694.00,564.47)(1.549,0.504){11}{\rule{1.087pt}{0.121pt}}
\multiput(694.00,563.27)(17.743,7.000){2}{\rule{0.544pt}{0.350pt}}
\multiput(714.00,571.47)(1.266,0.504){13}{\rule{0.919pt}{0.121pt}}
\multiput(714.00,570.27)(17.093,8.000){2}{\rule{0.459pt}{0.350pt}}
\multiput(733.00,579.47)(1.266,0.504){13}{\rule{0.919pt}{0.121pt}}
\multiput(733.00,578.27)(17.093,8.000){2}{\rule{0.459pt}{0.350pt}}
\multiput(752.00,587.47)(1.469,0.504){11}{\rule{1.038pt}{0.121pt}}
\multiput(752.00,586.27)(16.847,7.000){2}{\rule{0.519pt}{0.350pt}}
\multiput(771.00,594.47)(1.266,0.504){13}{\rule{0.919pt}{0.121pt}}
\multiput(771.00,593.27)(17.093,8.000){2}{\rule{0.459pt}{0.350pt}}
\multiput(790.00,602.47)(1.334,0.504){13}{\rule{0.962pt}{0.121pt}}
\multiput(790.00,601.27)(18.002,8.000){2}{\rule{0.481pt}{0.350pt}}
\multiput(810.00,610.47)(1.112,0.503){15}{\rule{0.826pt}{0.121pt}}
\multiput(810.00,609.27)(17.285,9.000){2}{\rule{0.413pt}{0.350pt}}
\multiput(829.00,619.47)(1.266,0.504){13}{\rule{0.919pt}{0.121pt}}
\multiput(829.00,618.27)(17.093,8.000){2}{\rule{0.459pt}{0.350pt}}
\multiput(848.00,627.47)(1.112,0.503){15}{\rule{0.826pt}{0.121pt}}
\multiput(848.00,626.27)(17.285,9.000){2}{\rule{0.413pt}{0.350pt}}
\multiput(867.00,636.47)(1.266,0.504){13}{\rule{0.919pt}{0.121pt}}
\multiput(867.00,635.27)(17.093,8.000){2}{\rule{0.459pt}{0.350pt}}
\multiput(886.00,644.47)(1.172,0.503){15}{\rule{0.865pt}{0.121pt}}
\multiput(886.00,643.27)(18.204,9.000){2}{\rule{0.433pt}{0.350pt}}
\multiput(906.00,653.47)(1.112,0.503){15}{\rule{0.826pt}{0.121pt}}
\multiput(906.00,652.27)(17.285,9.000){2}{\rule{0.413pt}{0.350pt}}
\multiput(925.00,662.47)(1.112,0.503){15}{\rule{0.826pt}{0.121pt}}
\multiput(925.00,661.27)(17.285,9.000){2}{\rule{0.413pt}{0.350pt}}
\multiput(944.00,671.47)(1.112,0.503){15}{\rule{0.826pt}{0.121pt}}
\multiput(944.00,670.27)(17.285,9.000){2}{\rule{0.413pt}{0.350pt}}
\multiput(963.00,680.47)(1.112,0.503){15}{\rule{0.826pt}{0.121pt}}
\multiput(963.00,679.27)(17.285,9.000){2}{\rule{0.413pt}{0.350pt}}
\multiput(982.00,689.48)(1.046,0.503){17}{\rule{0.787pt}{0.121pt}}
\multiput(982.00,688.27)(18.366,10.000){2}{\rule{0.394pt}{0.350pt}}
\multiput(1002.00,699.47)(1.112,0.503){15}{\rule{0.826pt}{0.121pt}}
\multiput(1002.00,698.27)(17.285,9.000){2}{\rule{0.413pt}{0.350pt}}
\multiput(1021.00,708.48)(0.992,0.503){17}{\rule{0.752pt}{0.121pt}}
\multiput(1021.00,707.27)(17.438,10.000){2}{\rule{0.376pt}{0.350pt}}
\multiput(1040.00,718.48)(0.992,0.503){17}{\rule{0.752pt}{0.121pt}}
\multiput(1040.00,717.27)(17.438,10.000){2}{\rule{0.376pt}{0.350pt}}
\multiput(1059.00,728.48)(0.992,0.503){17}{\rule{0.752pt}{0.121pt}}
\multiput(1059.00,727.27)(17.438,10.000){2}{\rule{0.376pt}{0.350pt}}
\multiput(1078.00,738.48)(1.046,0.503){17}{\rule{0.787pt}{0.121pt}}
\multiput(1078.00,737.27)(18.366,10.000){2}{\rule{0.394pt}{0.350pt}}
\multiput(1098.00,748.48)(0.992,0.503){17}{\rule{0.752pt}{0.121pt}}
\multiput(1098.00,747.27)(17.438,10.000){2}{\rule{0.376pt}{0.350pt}}
\multiput(1117.00,758.48)(0.896,0.502){19}{\rule{0.692pt}{0.121pt}}
\multiput(1117.00,757.27)(17.564,11.000){2}{\rule{0.346pt}{0.350pt}}
\end{picture}
\vspace{0.4cm}
\caption{Illustrative plot for changes of two $D^0$-$\bar{D}^0$ 
mixing observables with proper time $t$, where (a) $x^{~}_D = 
y^{~}_D = 0.05$, $\phi_D =\pi/4$ (corresponding to the solid curves)
and (b) $x^{~}_D =0.05$, $y^{~}_D =0$,
$\phi_D =\pi/2$ (corresponding to the dark solid curves) have been taken.}
\end{figure}

Finally it is worth pointing out that a comparison of the interference terms in 
(2.7) with $T^{\pm}_{\rm int}$ in (2.4) can provide a model-independent
constraint on the strong phase shift $\delta_{K\pi}$.

\section{Typical signals of $CP$ violation}
\setcounter{equation}{0}

It is expected that $CP$ violation induced by $D^0$-$\bar{D}^0$ mixing 
(i.e., $\Delta_D$) can manifest itself, apparently or indirectly, in all neutral $D$-meson
decay modes. But this effect should be negligibly small in most cases.
A constraint on $\Delta_D$ is possible through measuring the semileptonic 
decays of either incoherent or coherent $D^0$ and $\bar{D}^0$ mesons. For example,
\begin{equation}
\frac{{\cal R} (\bar{D}^0_{\rm phys}\rightarrow K^-l^+\nu^{~}_l) ~ - ~ 
{\cal R} (D^0_{\rm phys}\rightarrow K^+l^-\bar{\nu}^{~}_l)}
{{\cal R} (\bar{D}^0_{\rm phys}\rightarrow K^-l^+\nu^{~}_l) ~ + ~ 
{\cal R} (D^0_{\rm phys}\rightarrow K^+l^-\bar{\nu}^{~}_l)} \; = \; \Delta_D \; 
\end{equation}
and
\begin{equation}
\frac{{\cal R} (K^-l^+\nu^{~}_l; K^-l^+\nu^{~}_l)_C ~ - ~ 
{\cal R} (K^+l^-\bar{\nu}^{~}_l; K^+l^-\bar{\nu}^{~}_l)_C}
{{\cal R} (K^-l^+\nu^{~}_l; K^-l^+\nu^{~}_l)_C ~ + ~ 
{\cal R} (K^+l^-\bar{\nu}^{~}_l; K^+l^-\bar{\nu}^{~}_l)_C} \; = \; \Delta_D \; 
\end{equation}
for both $C$-odd and $C$-even cases.
In the following we shall pay main attention to the direct and indirect $CP$
asymmetries in some nonleptonic $D$ transitions, where $\Delta_D =0$ will
be assumed.

\begin{center}
{\large\bf A. ~ Time-integrated measurements}
\end{center}

For neutral $D$ mesons decaying to a hadronic $CP$ eigenstate $f$, the observables
of direct $CP$ violation in the decay amplitude and indirect $CP$ violation from
the interplay of decay and $D^0$-$\bar{D}^0$ mixing are expressed as
\begin{equation}
{\cal A}_{\rm dir} \; \equiv \; \frac{1 - |\rho^{~}_f|^2}{1 + |\rho^{~}_f|^2}
\; , ~~~~~~~~
{\cal A}_{\rm ind} \; \equiv \; \frac{-2 ~ {\rm Im} \left (e^{{\rm i}\phi_D} \rho^{~}_f
\right )}{1 + |\rho^{~}_f|^2} \; ,
\end{equation}
where $\rho^{~}_f \equiv \langle f|{\cal H}_{\rm eff}|\bar{D}^0\rangle /
\langle f|{\cal H}_{\rm eff}|D^0\rangle$. In the time-integrated measurements,
the following $CP$ asymmetries can be used to probe ${\cal A}_{\rm dir}$ and
${\cal A}_{\rm ind}$:

(a) For incoherent decays of $D^0$ and $\bar{D}^0$ mesons, we have
\begin{equation}
\frac{{\cal R} (D^0_{\rm phys}\rightarrow f) ~ - ~ 
{\cal R} (\bar{D}^0_{\rm phys}\rightarrow f)}{{\cal R} (D^0_{\rm phys}\rightarrow f)
~ + ~ {\cal R} (\bar{D}^0_{\rm phys}\rightarrow f)} \; \approx \;
{\cal A}_{\rm dir} ~ + ~ x^{~}_D {\cal A}_{\rm ind} \; .
\end{equation}
Note that a cancellation between ${\cal A}_{\rm dir}$ and $x^{~}_D {\cal A}_{\rm ind}$
may take place if they have the opposite signs, leading the above $CP$ asymmetry
to a negligibly small value.

(b) For coherent decays of $D^0\bar{D}^0$ pairs at the $\psi (3.77)$ or $\psi (4.16)$
resonance, one can get
\begin{equation}
\frac{{\cal R} (l^-X^+; f)_C ~ - ~ {\cal R} (l^+X^-; f)_C}
{{\cal R} (l^-X^+; f)_C ~ + ~ {\cal R} (l^+X^-; f)_C} \; \approx \;
\left \{ \begin{array}{ll}
{\cal A}_{\rm dir} & (C {\rm -odd}) \\
{\cal A}_{\rm dir} + 2 x^{~}_D {\cal A}_{\rm ind} ~~ & (C {\rm -even}) 
\end{array} \right . \; .
\end{equation}
Clearly it is possible to distinguish between direct and indirect $CP$-violating 
signals, if the magnitude of ${\cal A}_{\rm dir}$ is comparable with that of
$x^{~}_D {\cal A}_{\rm ind}$. 

(c) At the $\psi (4.16)$ resonance there may be a type of $CP$ violation arising
from the $CP$-forbidden decay channels. For example,
\begin{equation}
\frac{{\cal R} (f; f)_{C{\rm -odd}}}{{\cal R} (f; f)_{C{\rm -even}}}
\; \approx \; r^{~}_D \left ({\cal A}^2_{\rm dir} ~ + ~ {\cal A}^2_{\rm ind} \right ) \; ,
\end{equation}
where we have assumed ${\cal A}_{\rm dir} < 10\%$ and ${\cal A}_{\rm ind} < 10\%$.
Such a $CP$-violating signal is in principle interesting, but measuring it
might be very difficult due to the smallness of $r^{~}_D$.

A more special case is associated with $D^0 / \bar{D}^0 \rightarrow K_{S,L} +
\pi^0$, where $CP$ violation in the decay amplitude or that from $K^0$-$\bar{K}^0$
mixing can be neglected. We find, on the $\psi (3.77)$ resonance (i.e., $C$-odd), that
\begin{equation}
\frac{{\cal R} (K_S \pi^0; K_S \pi^0)}{{\cal R} (K_S \pi^0; K_L \pi^0)}
\; \approx \; \frac{{\cal R} (K_L \pi^0; K_L \pi^0)}{{\cal R} (K_S \pi^0;
K_L \pi^0)} \; \approx \; r^{~}_D \sin^2\phi_D \; .
\end{equation}
If $r^{~}_D$ were close to its experimental upper bound and $\phi_D$
were enhanced by new physics, this signal could be measured at a $\tau$-charm
factory
\footnote{In contrast with (3.7), there may be a similar $CP$-violating signal 
for $B_d$ decays to $K_S X_c$ and $K_L X_c$ on the $\Upsilon (4S)$ resonance,
where $X_c = J/\psi$, $\psi^{\prime}$, $\eta_c$, $\eta_c^{\prime}$, etc
\cite{XingShizuoka97}. This signal is expected to be of $O(10\%)$ due to 
the large $B^0_d$-$\bar{B}^0_d$ mixing rate ($x^{~}_B \approx 0.7$) and significant 
$CP$-violating phase ($\phi_B \sim 19^{\circ} - 70^{\circ}$) in the standard
model, thus it should be detectable at the forthcoming $B$-meson factories.}.

Now we turn our attention to $CP$ violation in neutral $D$ decays to non-$CP$
eigenstates. Again $D^0 / \bar{D}^0 \rightarrow K^{\pm}\pi^{\mp}$ can
be taken as a good example for illustration. It is expected that only
indirect $CP$ violation appears in these four decay modes. We denote
the signals as follows:
\begin{eqnarray}
{\cal A}_{K\pi} & \equiv & \sqrt{R_{\rm\scriptstyle DCSD}} \sin\phi_D
\left (y^{~}_D \sin\delta_{K\pi} ~ - ~ x^{~}_D \cos\delta_{K\pi} \right ) \; ,
\nonumber \\
{\cal A}^{\prime}_{K\pi} & \equiv & \sqrt{R_{\rm\scriptstyle DCSD}} \sin\phi_D
\left (y^{~}_D \sin\delta_{K\pi} ~ + ~ x^{~}_D \cos\delta_{K\pi} \right ) \; ,
\end{eqnarray}
where $\phi_D$ and $\delta_{K\pi}$ have been defined before. 
If $|y^{~}_D| \ll |x^{~}_D|$, as anticipated in some non-standard models 
\cite{Golowich97,Pakvasa96Wol95Nir95}, 
we arrive at ${\cal A}^{\prime}_{K\pi} \approx - {\cal A}_{K\pi}$. 
For incoherent $D$-meson decays, ${\cal A}_{K\pi}$ can be measured from the
decay-rate asymmetry
\begin{equation}
\frac{{\cal R} (\bar{D}^0_{\rm phys} \rightarrow K^+\pi^-) ~ - ~
{\cal R} (D^0_{\rm phys} \rightarrow K^-\pi^+)}
{{\cal R} (\bar{D}^0_{\rm phys} \rightarrow K^+\pi^-) ~ + ~
{\cal R} (D^0_{\rm phys} \rightarrow K^-\pi^+)} \; \approx \; {\cal A}_{K\pi} \; .
\end{equation}
$CP$-violating signals in coherent $D^0\bar{D}^0$ decays to the final
states $(l^{\pm}X^{\mp})_D ~ (K^{\pm}\pi^{\mp})_{\bar D}$
and $(K^{\pm}\pi^{\mp})_D ~ (K^{\mp}\pi^{\pm})_{\bar D}$ are listed in
Table 2, where the $C$-odd case (associated with vanishing $CP$ asymmetries)
is not included. 
\begin{center}
{Table 2: $CP$-violating effects in typical coherent $D^0\bar{D}^0$ decays
at the $\psi (4.16)$ resonance \cite{Xing97,Xing96}.} 
\begin{tabular}{cc} \\ \hline\hline \\ 
~~~~~~~ Observable ~~~~~~~	& ~~~~~~~ Signal ~ ($C$-even) ~~~~~~~ \\ \\ \hline \\
$\displaystyle \frac{{\cal R} (l^+ X^-; K^+ \pi^-)_C
~ - ~ {\cal R} (l^- X^+; K^- \pi^+)_C}
{{\cal R} (l^+ X^-; K^+\pi^-)_C ~ + ~ {\cal R} (l^-X^+; K^-\pi^+)_C}$ 	
& $ 2 {\cal A}_{K\pi}$ 	\\ \\ \hline \\
$\displaystyle \frac{{\cal R} (l^-X^+; K^+\pi^-)_C}
{{\cal R} (l^-X^+; K^-\pi^+)_C} ~ - ~
\frac{{\cal R} (l^+X^-; K^-\pi^+)_C}{{\cal R} (l^+X^-; K^+\pi^-)_C}$
& $4 {\cal A}^{\prime}_{K\pi}$ \\ \\ \hline \\
$\displaystyle \frac{{\cal R} (K^+\pi^-; K^+\pi^-)_C}
{{\cal R} (K^-\pi^+; K^+\pi^-)_C} ~ - ~
\frac{{\cal R} (K^-\pi^+; K^-\pi^+)_C}{{\cal R} (K^-\pi^+; K^+\pi^-)_C}$
& $8 {\cal A}^{\prime}_{K\pi}$ \\ \\ \hline \hline
\end{tabular}
\end{center}

We see from Table 2 that it is possible to measure (or constrain) ${\cal A}_{K\pi}$
and ${\cal A}^{\prime}_{K\pi}$ on the $\psi (4.16)$ resonance with
$C$-even $D^0\bar{D}^0$ events at a $\tau$-charm factory.

\begin{center}
{\large\bf B. ~ Time-dependent measurements}
\end{center}

For coherent $D^0\bar{D}^0$ decays at the $\psi (3.77)$ and
$\psi (4.16)$ resonances, to measure the time dependence of a joint decay mode requires
asymmetric $e^+e^-$ collisions, like the case of an asymmetric $B$-meson factory.
A brief discussion about this possibility can be found in Appendix A of
Ref. \cite{Xing97}.
For incoherent neutral $D$-meson decays to a $CP$ eigenstate $f$, one can get the
time-dependent $CP$ asymmetry 
\begin{equation}
\frac{{\cal R} [D^0(t)\rightarrow f] ~ - ~ {\cal R} [\bar{D}^0(t)\rightarrow f]}
{{\cal R} [D^0(t)\rightarrow f] ~ + ~ {\cal R} [\bar{D}^0(t)\rightarrow f]}
\; \approx \; {\cal A}_{\rm dir} ~ + ~ x^{~}_D {\cal A}_{\rm ind} \left (\Gamma t
\right ) \; ,
\end{equation}
where ${\cal A}_{\rm dir}$ and ${\cal A}_{\rm ind}$ have been defined before.

Taking $D^0 /\bar{D}^0\rightarrow K_{S,L} + \pi^0$ for example, we obtain
\begin{eqnarray}
\frac{{\cal R} [D^0(t)\rightarrow K_S \pi^0] ~ - ~ {\cal R} [\bar{D}^0(t)\rightarrow 
K_S \pi^0]}{{\cal R} [D^0(t)\rightarrow K_S \pi^0] ~ + ~ 
{\cal R} [\bar{D}^0(t)\rightarrow K_S \pi^0]}
& \approx & -2 {\rm Re} \epsilon^{~}_K ~ + ~ x^{~}_D \sin\phi_D \left (\Gamma t
\right ) \; , \nonumber \\
\frac{{\cal R} [D^0(t)\rightarrow K_L \pi^0] ~ - ~ {\cal R} [\bar{D}^0(t)\rightarrow 
K_L \pi^0]}{{\cal R} [D^0(t)\rightarrow K_L \pi^0] ~ + ~ 
{\cal R} [\bar{D}^0(t)\rightarrow K_L \pi^0]}
& \approx & -2 {\rm Re} \epsilon^{~}_K ~ - ~ x^{~}_D \sin\phi_D \left (\Gamma t
\right ) \; 
\end{eqnarray}
in the assumption of $\Delta_D =0$. Here ${\rm Re}\epsilon^{~}_K \approx
1.6\times 10^{-3}$, signifying the $CP$ asymmetry induced by $K^0$-$\bar{K}^0$
mixing, cannot be neglected \cite{Xing95}. Even if the $x^{~}_D \sin\phi_D$ term is 
vanishingly small, the effect of ${\rm Re}\epsilon^{~}_K$ is still detectable
from the above decay modes. For the purpose of illustration, we take $x^{~}_D =0.01$
and $\phi_D =0.1$ to plot changes of the $CP$ asymmetries (3.11) with
proper time $t$ in Fig. 3.
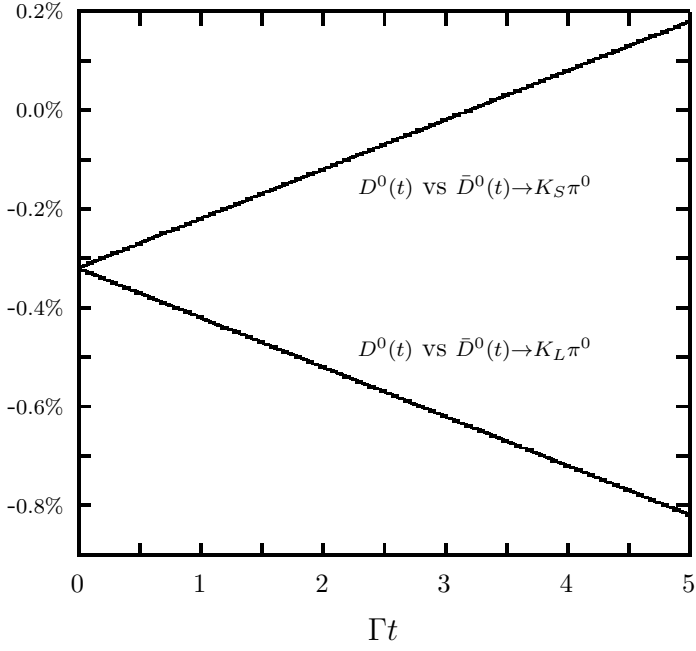
\begin{figure}
\setlength{\unitlength}{0.240900pt}
\ifx\plotpoint\undefined\newsavebox{\plotpoint}\fi
\sbox{\plotpoint}{\rule[-0.500pt]{1.000pt}{1.000pt}}%
\begin{picture}(1200,990)(-350,0)
\font\gnuplot=cmr10 at 10pt
\gnuplot
\sbox{\plotpoint}{\rule[-0.500pt]{1.000pt}{1.000pt}}%
\put(176.0,113.0){\rule[-0.500pt]{1.000pt}{205.729pt}}
\put(176.0,113.0){\rule[-0.500pt]{4.818pt}{1.000pt}}
\put(1116.0,113.0){\rule[-0.500pt]{4.818pt}{1.000pt}}
\put(176.0,191.0){\rule[-0.500pt]{4.818pt}{1.000pt}}
\put(154,191){\makebox(0,0)[r]{-$\scriptstyle 0.8\%$}}
\put(1116.0,191.0){\rule[-0.500pt]{4.818pt}{1.000pt}}
\put(176.0,268.0){\rule[-0.500pt]{4.818pt}{1.000pt}}
\put(1116.0,268.0){\rule[-0.500pt]{4.818pt}{1.000pt}}
\put(176.0,346.0){\rule[-0.500pt]{4.818pt}{1.000pt}}
\put(154,346){\makebox(0,0)[r]{-$\scriptstyle 0.6\%$}}
\put(1116.0,346.0){\rule[-0.500pt]{4.818pt}{1.000pt}}
\put(176.0,424.0){\rule[-0.500pt]{4.818pt}{1.000pt}}
\put(1116.0,424.0){\rule[-0.500pt]{4.818pt}{1.000pt}}
\put(176.0,501.0){\rule[-0.500pt]{4.818pt}{1.000pt}}
\put(154,501){\makebox(0,0)[r]{-$\scriptstyle 0.4\%$}}
\put(1116.0,501.0){\rule[-0.500pt]{4.818pt}{1.000pt}}
\put(176.0,579.0){\rule[-0.500pt]{4.818pt}{1.000pt}}
\put(1116.0,579.0){\rule[-0.500pt]{4.818pt}{1.000pt}}
\put(176.0,656.0){\rule[-0.500pt]{4.818pt}{1.000pt}}
\put(154,656){\makebox(0,0)[r]{-$\scriptstyle 0.2\%$}}
\put(1116.0,656.0){\rule[-0.500pt]{4.818pt}{1.000pt}}
\put(176.0,734.0){\rule[-0.500pt]{4.818pt}{1.000pt}}
\put(1116.0,734.0){\rule[-0.500pt]{4.818pt}{1.000pt}}
\put(176.0,812.0){\rule[-0.500pt]{4.818pt}{1.000pt}}
\put(154,812){\makebox(0,0)[r]{$\scriptstyle 0.0\%$}}
\put(1116.0,812.0){\rule[-0.500pt]{4.818pt}{1.000pt}}
\put(176.0,889.0){\rule[-0.500pt]{4.818pt}{1.000pt}}
\put(1116.0,889.0){\rule[-0.500pt]{4.818pt}{1.000pt}}
\put(176.0,967.0){\rule[-0.500pt]{4.818pt}{1.000pt}}
\put(154,967){\makebox(0,0)[r]{$\scriptstyle 0.2\%$}}
\put(1116.0,967.0){\rule[-0.500pt]{4.818pt}{1.000pt}}
\put(176.0,113.0){\rule[-0.500pt]{1.000pt}{4.818pt}}
\put(176,68){\makebox(0,0){0}}
\put(176.0,947.0){\rule[-0.500pt]{1.000pt}{4.818pt}}
\put(272.0,113.0){\rule[-0.500pt]{1.000pt}{4.818pt}}
\put(272.0,947.0){\rule[-0.500pt]{1.000pt}{4.818pt}}
\put(368.0,113.0){\rule[-0.500pt]{1.000pt}{4.818pt}}
\put(368,68){\makebox(0,0){1}}
\put(368.0,947.0){\rule[-0.500pt]{1.000pt}{4.818pt}}
\put(464.0,113.0){\rule[-0.500pt]{1.000pt}{4.818pt}}
\put(464.0,947.0){\rule[-0.500pt]{1.000pt}{4.818pt}}
\put(560.0,113.0){\rule[-0.500pt]{1.000pt}{4.818pt}}
\put(560,68){\makebox(0,0){2}}
\put(560.0,947.0){\rule[-0.500pt]{1.000pt}{4.818pt}}
\put(656.0,113.0){\rule[-0.500pt]{1.000pt}{4.818pt}}
\put(656.0,947.0){\rule[-0.500pt]{1.000pt}{4.818pt}}
\put(752.0,113.0){\rule[-0.500pt]{1.000pt}{4.818pt}}
\put(752,68){\makebox(0,0){3}}
\put(752.0,947.0){\rule[-0.500pt]{1.000pt}{4.818pt}}
\put(848.0,113.0){\rule[-0.500pt]{1.000pt}{4.818pt}}
\put(848.0,947.0){\rule[-0.500pt]{1.000pt}{4.818pt}}
\put(944.0,113.0){\rule[-0.500pt]{1.000pt}{4.818pt}}
\put(944,68){\makebox(0,0){4}}
\put(944.0,947.0){\rule[-0.500pt]{1.000pt}{4.818pt}}
\put(1040.0,113.0){\rule[-0.500pt]{1.000pt}{4.818pt}}
\put(1040.0,947.0){\rule[-0.500pt]{1.000pt}{4.818pt}}
\put(1136.0,113.0){\rule[-0.500pt]{1.000pt}{4.818pt}}
\put(1136,68){\makebox(0,0){5}}
\put(1136.0,947.0){\rule[-0.500pt]{1.000pt}{4.818pt}}
\put(176.0,113.0){\rule[-0.500pt]{231.264pt}{1.000pt}}
\put(1136.0,113.0){\rule[-0.500pt]{1.000pt}{205.729pt}}
\put(176.0,967.0){\rule[-0.500pt]{231.264pt}{1.000pt}}
\put(656,-3){\makebox(0,0){$\Gamma t$}}
\put(800,690){\makebox(0,0){$\scriptstyle D^0(t)$ vs $\scriptstyle 
	     \bar{D}^0(t)\rightarrow K_S \pi^0$}}
\put(800,435){\makebox(0,0){$\scriptstyle D^0(t)$ vs $\scriptstyle
	     \bar{D}^0(t)\rightarrow K_L \pi^0$}}

\put(176.0,113.0){\rule[-0.500pt]{1.000pt}{205.729pt}}
\put(176,563){\usebox{\plotpoint}}
\multiput(176.00,564.83)(1.158,0.481){8}{\rule{2.625pt}{0.116pt}}
\multiput(176.00,560.92)(13.552,8.000){2}{\rule{1.313pt}{1.000pt}}
\multiput(195.00,572.83)(1.158,0.481){8}{\rule{2.625pt}{0.116pt}}
\multiput(195.00,568.92)(13.552,8.000){2}{\rule{1.313pt}{1.000pt}}
\multiput(214.00,580.83)(1.226,0.481){8}{\rule{2.750pt}{0.116pt}}
\multiput(214.00,576.92)(14.292,8.000){2}{\rule{1.375pt}{1.000pt}}
\multiput(234.00,588.84)(1.339,0.475){6}{\rule{2.964pt}{0.114pt}}
\multiput(234.00,584.92)(12.847,7.000){2}{\rule{1.482pt}{1.000pt}}
\multiput(253.00,595.83)(1.158,0.481){8}{\rule{2.625pt}{0.116pt}}
\multiput(253.00,591.92)(13.552,8.000){2}{\rule{1.313pt}{1.000pt}}
\multiput(272.00,603.83)(1.158,0.481){8}{\rule{2.625pt}{0.116pt}}
\multiput(272.00,599.92)(13.552,8.000){2}{\rule{1.313pt}{1.000pt}}
\multiput(291.00,611.83)(1.158,0.481){8}{\rule{2.625pt}{0.116pt}}
\multiput(291.00,607.92)(13.552,8.000){2}{\rule{1.313pt}{1.000pt}}
\multiput(310.00,619.84)(1.420,0.475){6}{\rule{3.107pt}{0.114pt}}
\multiput(310.00,615.92)(13.551,7.000){2}{\rule{1.554pt}{1.000pt}}
\multiput(330.00,626.83)(1.158,0.481){8}{\rule{2.625pt}{0.116pt}}
\multiput(330.00,622.92)(13.552,8.000){2}{\rule{1.313pt}{1.000pt}}
\multiput(349.00,634.83)(1.158,0.481){8}{\rule{2.625pt}{0.116pt}}
\multiput(349.00,630.92)(13.552,8.000){2}{\rule{1.313pt}{1.000pt}}
\multiput(368.00,642.83)(1.158,0.481){8}{\rule{2.625pt}{0.116pt}}
\multiput(368.00,638.92)(13.552,8.000){2}{\rule{1.313pt}{1.000pt}}
\multiput(387.00,650.84)(1.339,0.475){6}{\rule{2.964pt}{0.114pt}}
\multiput(387.00,646.92)(12.847,7.000){2}{\rule{1.482pt}{1.000pt}}
\multiput(406.00,657.83)(1.226,0.481){8}{\rule{2.750pt}{0.116pt}}
\multiput(406.00,653.92)(14.292,8.000){2}{\rule{1.375pt}{1.000pt}}
\multiput(426.00,665.83)(1.158,0.481){8}{\rule{2.625pt}{0.116pt}}
\multiput(426.00,661.92)(13.552,8.000){2}{\rule{1.313pt}{1.000pt}}
\multiput(445.00,673.83)(1.158,0.481){8}{\rule{2.625pt}{0.116pt}}
\multiput(445.00,669.92)(13.552,8.000){2}{\rule{1.313pt}{1.000pt}}
\multiput(464.00,681.84)(1.339,0.475){6}{\rule{2.964pt}{0.114pt}}
\multiput(464.00,677.92)(12.847,7.000){2}{\rule{1.482pt}{1.000pt}}
\multiput(483.00,688.83)(1.158,0.481){8}{\rule{2.625pt}{0.116pt}}
\multiput(483.00,684.92)(13.552,8.000){2}{\rule{1.313pt}{1.000pt}}
\multiput(502.00,696.83)(1.226,0.481){8}{\rule{2.750pt}{0.116pt}}
\multiput(502.00,692.92)(14.292,8.000){2}{\rule{1.375pt}{1.000pt}}
\multiput(522.00,704.83)(1.158,0.481){8}{\rule{2.625pt}{0.116pt}}
\multiput(522.00,700.92)(13.552,8.000){2}{\rule{1.313pt}{1.000pt}}
\multiput(541.00,712.84)(1.339,0.475){6}{\rule{2.964pt}{0.114pt}}
\multiput(541.00,708.92)(12.847,7.000){2}{\rule{1.482pt}{1.000pt}}
\multiput(560.00,719.83)(1.158,0.481){8}{\rule{2.625pt}{0.116pt}}
\multiput(560.00,715.92)(13.552,8.000){2}{\rule{1.313pt}{1.000pt}}
\multiput(579.00,727.83)(1.158,0.481){8}{\rule{2.625pt}{0.116pt}}
\multiput(579.00,723.92)(13.552,8.000){2}{\rule{1.313pt}{1.000pt}}
\multiput(598.00,735.83)(1.226,0.481){8}{\rule{2.750pt}{0.116pt}}
\multiput(598.00,731.92)(14.292,8.000){2}{\rule{1.375pt}{1.000pt}}
\multiput(618.00,743.84)(1.339,0.475){6}{\rule{2.964pt}{0.114pt}}
\multiput(618.00,739.92)(12.847,7.000){2}{\rule{1.482pt}{1.000pt}}
\multiput(637.00,750.83)(1.158,0.481){8}{\rule{2.625pt}{0.116pt}}
\multiput(637.00,746.92)(13.552,8.000){2}{\rule{1.313pt}{1.000pt}}
\multiput(656.00,758.83)(1.158,0.481){8}{\rule{2.625pt}{0.116pt}}
\multiput(656.00,754.92)(13.552,8.000){2}{\rule{1.313pt}{1.000pt}}
\multiput(675.00,766.83)(1.158,0.481){8}{\rule{2.625pt}{0.116pt}}
\multiput(675.00,762.92)(13.552,8.000){2}{\rule{1.313pt}{1.000pt}}
\multiput(694.00,774.84)(1.420,0.475){6}{\rule{3.107pt}{0.114pt}}
\multiput(694.00,770.92)(13.551,7.000){2}{\rule{1.554pt}{1.000pt}}
\multiput(714.00,781.83)(1.158,0.481){8}{\rule{2.625pt}{0.116pt}}
\multiput(714.00,777.92)(13.552,8.000){2}{\rule{1.313pt}{1.000pt}}
\multiput(733.00,789.83)(1.158,0.481){8}{\rule{2.625pt}{0.116pt}}
\multiput(733.00,785.92)(13.552,8.000){2}{\rule{1.313pt}{1.000pt}}
\multiput(752.00,797.83)(1.158,0.481){8}{\rule{2.625pt}{0.116pt}}
\multiput(752.00,793.92)(13.552,8.000){2}{\rule{1.313pt}{1.000pt}}
\multiput(771.00,805.84)(1.339,0.475){6}{\rule{2.964pt}{0.114pt}}
\multiput(771.00,801.92)(12.847,7.000){2}{\rule{1.482pt}{1.000pt}}
\multiput(790.00,812.83)(1.226,0.481){8}{\rule{2.750pt}{0.116pt}}
\multiput(790.00,808.92)(14.292,8.000){2}{\rule{1.375pt}{1.000pt}}
\multiput(810.00,820.83)(1.158,0.481){8}{\rule{2.625pt}{0.116pt}}
\multiput(810.00,816.92)(13.552,8.000){2}{\rule{1.313pt}{1.000pt}}
\multiput(829.00,828.83)(1.158,0.481){8}{\rule{2.625pt}{0.116pt}}
\multiput(829.00,824.92)(13.552,8.000){2}{\rule{1.313pt}{1.000pt}}
\multiput(848.00,836.84)(1.339,0.475){6}{\rule{2.964pt}{0.114pt}}
\multiput(848.00,832.92)(12.847,7.000){2}{\rule{1.482pt}{1.000pt}}
\multiput(867.00,843.83)(1.158,0.481){8}{\rule{2.625pt}{0.116pt}}
\multiput(867.00,839.92)(13.552,8.000){2}{\rule{1.313pt}{1.000pt}}
\multiput(886.00,851.83)(1.226,0.481){8}{\rule{2.750pt}{0.116pt}}
\multiput(886.00,847.92)(14.292,8.000){2}{\rule{1.375pt}{1.000pt}}
\multiput(906.00,859.83)(1.158,0.481){8}{\rule{2.625pt}{0.116pt}}
\multiput(906.00,855.92)(13.552,8.000){2}{\rule{1.313pt}{1.000pt}}
\multiput(925.00,867.84)(1.339,0.475){6}{\rule{2.964pt}{0.114pt}}
\multiput(925.00,863.92)(12.847,7.000){2}{\rule{1.482pt}{1.000pt}}
\multiput(944.00,874.83)(1.158,0.481){8}{\rule{2.625pt}{0.116pt}}
\multiput(944.00,870.92)(13.552,8.000){2}{\rule{1.313pt}{1.000pt}}
\multiput(963.00,882.83)(1.158,0.481){8}{\rule{2.625pt}{0.116pt}}
\multiput(963.00,878.92)(13.552,8.000){2}{\rule{1.313pt}{1.000pt}}
\multiput(982.00,890.83)(1.226,0.481){8}{\rule{2.750pt}{0.116pt}}
\multiput(982.00,886.92)(14.292,8.000){2}{\rule{1.375pt}{1.000pt}}
\multiput(1002.00,898.84)(1.339,0.475){6}{\rule{2.964pt}{0.114pt}}
\multiput(1002.00,894.92)(12.847,7.000){2}{\rule{1.482pt}{1.000pt}}
\multiput(1021.00,905.83)(1.158,0.481){8}{\rule{2.625pt}{0.116pt}}
\multiput(1021.00,901.92)(13.552,8.000){2}{\rule{1.313pt}{1.000pt}}
\multiput(1040.00,913.83)(1.158,0.481){8}{\rule{2.625pt}{0.116pt}}
\multiput(1040.00,909.92)(13.552,8.000){2}{\rule{1.313pt}{1.000pt}}
\multiput(1059.00,921.83)(1.158,0.481){8}{\rule{2.625pt}{0.116pt}}
\multiput(1059.00,917.92)(13.552,8.000){2}{\rule{1.313pt}{1.000pt}}
\multiput(1078.00,929.84)(1.420,0.475){6}{\rule{3.107pt}{0.114pt}}
\multiput(1078.00,925.92)(13.551,7.000){2}{\rule{1.554pt}{1.000pt}}
\multiput(1098.00,936.83)(1.158,0.481){8}{\rule{2.625pt}{0.116pt}}
\multiput(1098.00,932.92)(13.552,8.000){2}{\rule{1.313pt}{1.000pt}}
\multiput(1117.00,944.83)(1.158,0.481){8}{\rule{2.625pt}{0.116pt}}
\multiput(1117.00,940.92)(13.552,8.000){2}{\rule{1.313pt}{1.000pt}}
\put(176,563){\usebox{\plotpoint}}
\multiput(176.00,560.69)(1.339,-0.475){6}{\rule{2.964pt}{0.114pt}}
\multiput(176.00,560.92)(12.847,-7.000){2}{\rule{1.482pt}{1.000pt}}
\multiput(195.00,553.68)(1.158,-0.481){8}{\rule{2.625pt}{0.116pt}}
\multiput(195.00,553.92)(13.552,-8.000){2}{\rule{1.313pt}{1.000pt}}
\multiput(214.00,545.68)(1.226,-0.481){8}{\rule{2.750pt}{0.116pt}}
\multiput(214.00,545.92)(14.292,-8.000){2}{\rule{1.375pt}{1.000pt}}
\multiput(234.00,537.68)(1.158,-0.481){8}{\rule{2.625pt}{0.116pt}}
\multiput(234.00,537.92)(13.552,-8.000){2}{\rule{1.313pt}{1.000pt}}
\multiput(253.00,529.69)(1.339,-0.475){6}{\rule{2.964pt}{0.114pt}}
\multiput(253.00,529.92)(12.847,-7.000){2}{\rule{1.482pt}{1.000pt}}
\multiput(272.00,522.68)(1.158,-0.481){8}{\rule{2.625pt}{0.116pt}}
\multiput(272.00,522.92)(13.552,-8.000){2}{\rule{1.313pt}{1.000pt}}
\multiput(291.00,514.68)(1.158,-0.481){8}{\rule{2.625pt}{0.116pt}}
\multiput(291.00,514.92)(13.552,-8.000){2}{\rule{1.313pt}{1.000pt}}
\multiput(310.00,506.68)(1.226,-0.481){8}{\rule{2.750pt}{0.116pt}}
\multiput(310.00,506.92)(14.292,-8.000){2}{\rule{1.375pt}{1.000pt}}
\multiput(330.00,498.69)(1.339,-0.475){6}{\rule{2.964pt}{0.114pt}}
\multiput(330.00,498.92)(12.847,-7.000){2}{\rule{1.482pt}{1.000pt}}
\multiput(349.00,491.68)(1.158,-0.481){8}{\rule{2.625pt}{0.116pt}}
\multiput(349.00,491.92)(13.552,-8.000){2}{\rule{1.313pt}{1.000pt}}
\multiput(368.00,483.68)(1.158,-0.481){8}{\rule{2.625pt}{0.116pt}}
\multiput(368.00,483.92)(13.552,-8.000){2}{\rule{1.313pt}{1.000pt}}
\multiput(387.00,475.68)(1.158,-0.481){8}{\rule{2.625pt}{0.116pt}}
\multiput(387.00,475.92)(13.552,-8.000){2}{\rule{1.313pt}{1.000pt}}
\multiput(406.00,467.69)(1.420,-0.475){6}{\rule{3.107pt}{0.114pt}}
\multiput(406.00,467.92)(13.551,-7.000){2}{\rule{1.554pt}{1.000pt}}
\multiput(426.00,460.68)(1.158,-0.481){8}{\rule{2.625pt}{0.116pt}}
\multiput(426.00,460.92)(13.552,-8.000){2}{\rule{1.313pt}{1.000pt}}
\multiput(445.00,452.68)(1.158,-0.481){8}{\rule{2.625pt}{0.116pt}}
\multiput(445.00,452.92)(13.552,-8.000){2}{\rule{1.313pt}{1.000pt}}
\multiput(464.00,444.68)(1.158,-0.481){8}{\rule{2.625pt}{0.116pt}}
\multiput(464.00,444.92)(13.552,-8.000){2}{\rule{1.313pt}{1.000pt}}
\multiput(483.00,436.69)(1.339,-0.475){6}{\rule{2.964pt}{0.114pt}}
\multiput(483.00,436.92)(12.847,-7.000){2}{\rule{1.482pt}{1.000pt}}
\multiput(502.00,429.68)(1.226,-0.481){8}{\rule{2.750pt}{0.116pt}}
\multiput(502.00,429.92)(14.292,-8.000){2}{\rule{1.375pt}{1.000pt}}
\multiput(522.00,421.68)(1.158,-0.481){8}{\rule{2.625pt}{0.116pt}}
\multiput(522.00,421.92)(13.552,-8.000){2}{\rule{1.313pt}{1.000pt}}
\multiput(541.00,413.68)(1.158,-0.481){8}{\rule{2.625pt}{0.116pt}}
\multiput(541.00,413.92)(13.552,-8.000){2}{\rule{1.313pt}{1.000pt}}
\multiput(560.00,405.69)(1.339,-0.475){6}{\rule{2.964pt}{0.114pt}}
\multiput(560.00,405.92)(12.847,-7.000){2}{\rule{1.482pt}{1.000pt}}
\multiput(579.00,398.68)(1.158,-0.481){8}{\rule{2.625pt}{0.116pt}}
\multiput(579.00,398.92)(13.552,-8.000){2}{\rule{1.313pt}{1.000pt}}
\multiput(598.00,390.68)(1.226,-0.481){8}{\rule{2.750pt}{0.116pt}}
\multiput(598.00,390.92)(14.292,-8.000){2}{\rule{1.375pt}{1.000pt}}
\multiput(618.00,382.68)(1.158,-0.481){8}{\rule{2.625pt}{0.116pt}}
\multiput(618.00,382.92)(13.552,-8.000){2}{\rule{1.313pt}{1.000pt}}
\multiput(637.00,374.69)(1.339,-0.475){6}{\rule{2.964pt}{0.114pt}}
\multiput(637.00,374.92)(12.847,-7.000){2}{\rule{1.482pt}{1.000pt}}
\multiput(656.00,367.68)(1.158,-0.481){8}{\rule{2.625pt}{0.116pt}}
\multiput(656.00,367.92)(13.552,-8.000){2}{\rule{1.313pt}{1.000pt}}
\multiput(675.00,359.68)(1.158,-0.481){8}{\rule{2.625pt}{0.116pt}}
\multiput(675.00,359.92)(13.552,-8.000){2}{\rule{1.313pt}{1.000pt}}
\multiput(694.00,351.68)(1.226,-0.481){8}{\rule{2.750pt}{0.116pt}}
\multiput(694.00,351.92)(14.292,-8.000){2}{\rule{1.375pt}{1.000pt}}
\multiput(714.00,343.69)(1.339,-0.475){6}{\rule{2.964pt}{0.114pt}}
\multiput(714.00,343.92)(12.847,-7.000){2}{\rule{1.482pt}{1.000pt}}
\multiput(733.00,336.68)(1.158,-0.481){8}{\rule{2.625pt}{0.116pt}}
\multiput(733.00,336.92)(13.552,-8.000){2}{\rule{1.313pt}{1.000pt}}
\multiput(752.00,328.68)(1.158,-0.481){8}{\rule{2.625pt}{0.116pt}}
\multiput(752.00,328.92)(13.552,-8.000){2}{\rule{1.313pt}{1.000pt}}
\multiput(771.00,320.68)(1.158,-0.481){8}{\rule{2.625pt}{0.116pt}}
\multiput(771.00,320.92)(13.552,-8.000){2}{\rule{1.313pt}{1.000pt}}
\multiput(790.00,312.69)(1.420,-0.475){6}{\rule{3.107pt}{0.114pt}}
\multiput(790.00,312.92)(13.551,-7.000){2}{\rule{1.554pt}{1.000pt}}
\multiput(810.00,305.68)(1.158,-0.481){8}{\rule{2.625pt}{0.116pt}}
\multiput(810.00,305.92)(13.552,-8.000){2}{\rule{1.313pt}{1.000pt}}
\multiput(829.00,297.68)(1.158,-0.481){8}{\rule{2.625pt}{0.116pt}}
\multiput(829.00,297.92)(13.552,-8.000){2}{\rule{1.313pt}{1.000pt}}
\multiput(848.00,289.68)(1.158,-0.481){8}{\rule{2.625pt}{0.116pt}}
\multiput(848.00,289.92)(13.552,-8.000){2}{\rule{1.313pt}{1.000pt}}
\multiput(867.00,281.69)(1.339,-0.475){6}{\rule{2.964pt}{0.114pt}}
\multiput(867.00,281.92)(12.847,-7.000){2}{\rule{1.482pt}{1.000pt}}
\multiput(886.00,274.68)(1.226,-0.481){8}{\rule{2.750pt}{0.116pt}}
\multiput(886.00,274.92)(14.292,-8.000){2}{\rule{1.375pt}{1.000pt}}
\multiput(906.00,266.68)(1.158,-0.481){8}{\rule{2.625pt}{0.116pt}}
\multiput(906.00,266.92)(13.552,-8.000){2}{\rule{1.313pt}{1.000pt}}
\multiput(925.00,258.68)(1.158,-0.481){8}{\rule{2.625pt}{0.116pt}}
\multiput(925.00,258.92)(13.552,-8.000){2}{\rule{1.313pt}{1.000pt}}
\multiput(944.00,250.69)(1.339,-0.475){6}{\rule{2.964pt}{0.114pt}}
\multiput(944.00,250.92)(12.847,-7.000){2}{\rule{1.482pt}{1.000pt}}
\multiput(963.00,243.68)(1.158,-0.481){8}{\rule{2.625pt}{0.116pt}}
\multiput(963.00,243.92)(13.552,-8.000){2}{\rule{1.313pt}{1.000pt}}
\multiput(982.00,235.68)(1.226,-0.481){8}{\rule{2.750pt}{0.116pt}}
\multiput(982.00,235.92)(14.292,-8.000){2}{\rule{1.375pt}{1.000pt}}
\multiput(1002.00,227.68)(1.158,-0.481){8}{\rule{2.625pt}{0.116pt}}
\multiput(1002.00,227.92)(13.552,-8.000){2}{\rule{1.313pt}{1.000pt}}
\multiput(1021.00,219.69)(1.339,-0.475){6}{\rule{2.964pt}{0.114pt}}
\multiput(1021.00,219.92)(12.847,-7.000){2}{\rule{1.482pt}{1.000pt}}
\multiput(1040.00,212.68)(1.158,-0.481){8}{\rule{2.625pt}{0.116pt}}
\multiput(1040.00,212.92)(13.552,-8.000){2}{\rule{1.313pt}{1.000pt}}
\multiput(1059.00,204.68)(1.158,-0.481){8}{\rule{2.625pt}{0.116pt}}
\multiput(1059.00,204.92)(13.552,-8.000){2}{\rule{1.313pt}{1.000pt}}
\multiput(1078.00,196.68)(1.226,-0.481){8}{\rule{2.750pt}{0.116pt}}
\multiput(1078.00,196.92)(14.292,-8.000){2}{\rule{1.375pt}{1.000pt}}
\multiput(1098.00,188.69)(1.339,-0.475){6}{\rule{2.964pt}{0.114pt}}
\multiput(1098.00,188.92)(12.847,-7.000){2}{\rule{1.482pt}{1.000pt}}
\multiput(1117.00,181.68)(1.158,-0.481){8}{\rule{2.625pt}{0.116pt}}
\multiput(1117.00,181.92)(13.552,-8.000){2}{\rule{1.313pt}{1.000pt}}
\end{picture}
\vspace{0.4cm}
\caption{Illustrative plot for changes of $CP$-violating asymmetries
with proper time $t$, where $x^{~}_D =0.01$ and 
$\phi_D =0.1$ have been taken.}
\end{figure}

Indirect $CP$ violation in neutral $D$-meson decays to hadronic non-$CP$ eigenstates
can be illustrated by taking $D^0 /\bar{D}^0\rightarrow K^{\pm}\pi^{\mp}$
for example. Assuming $\Delta_D =0$, we have
\begin{equation}
\frac{{\cal R} [\bar{D}^0(t)\rightarrow K^+\pi^-] ~ - ~ {\cal R} [D^0(t)\rightarrow 
K^-\pi^+]}{{\cal R} [\bar{D}^0(t)\rightarrow K^+\pi^-] ~ + ~ 
{\cal R} [D^0(t)\rightarrow K^-\pi^+]} \; \approx \; {\cal A}_{K\pi} \left (\Gamma t \right ) \; ,
\end{equation}
where ${\cal A}_{K\pi}$ has been given in Eq. (3.8).
Another $CP$ asymmetry reads
\begin{equation}
\frac{{\cal R} [D^0(t)\rightarrow K^+\pi^-] ~ - ~ {\cal R} [\bar{D}^0(t)\rightarrow 
K^- \pi^+]}{{\cal R} [D^0(t)\rightarrow K^+ \pi^-] ~ + ~ 
{\cal R} [\bar{D}^0(t)\rightarrow K^- \pi^+]}
\; \approx \; \frac{{\cal A}^{\prime}_{K\pi}}{N_{K\pi}} \left (\Gamma t \right ) \; 
\end{equation}
with
\begin{equation}
N_{K\pi} \; \equiv \; R_{\rm\scriptstyle DCSD} ~ + ~ \frac{T^+_{\rm int}
+ T^-_{\rm int}}{2} \left (\Gamma t\right ) ~ + ~ \frac{r^{~}_D}{2}
\left (\Gamma t\right )^2 \; .
\end{equation}
Here ${\cal A}^{\prime}_{K\pi}$ and $T^{\pm}_{\rm int}$ have been defined in 
Eqs. (3.8) and (2.4), respectively.
Obviously the asymmetry (3.13) may be large enough or even maximum in magnitude,
due to the smallness of $N_{K\pi}$ suppressed by DCSD and mixing effects.
To give one a numerical feeling, we plot changes of the $CP$ asymmetries (3.12) and (3.13) 
with proper time $t$ in Fig. 4 by taking $x^{~}_D =0.05$, $y^{~}_D =0$,
$\phi_D =\pi/2$, $\delta_{K\pi}=0$ and $R_{\rm\scriptstyle DCSD} =0.7\%$.
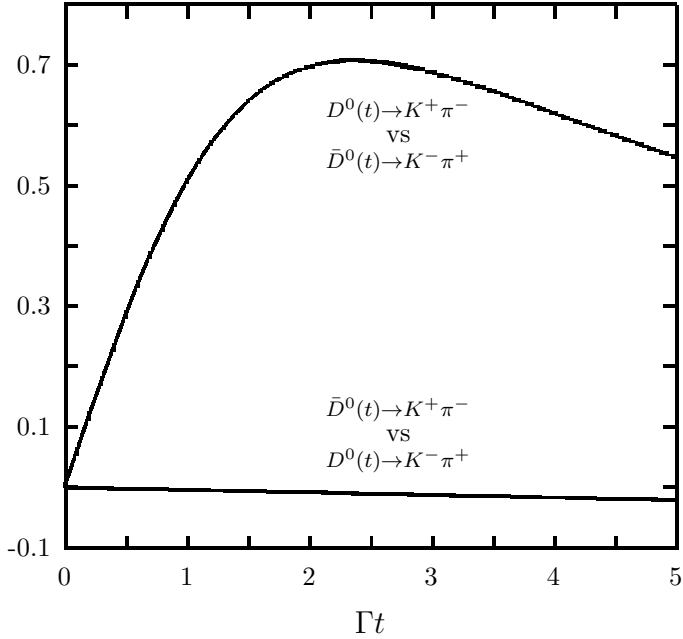
\begin{figure}
\setlength{\unitlength}{0.240900pt}
\ifx\plotpoint\undefined\newsavebox{\plotpoint}\fi
\sbox{\plotpoint}{\rule[-0.500pt]{1.000pt}{1.000pt}}%
\begin{picture}(1200,990)(-350,0)
\font\gnuplot=cmr10 at 10pt
\gnuplot
\sbox{\plotpoint}{\rule[-0.500pt]{1.000pt}{1.000pt}}%
\put(176.0,113.0){\rule[-0.500pt]{1.000pt}{205.729pt}}
\put(176.0,113.0){\rule[-0.500pt]{4.818pt}{1.000pt}}
\put(154,113){\makebox(0,0)[r]{-0.1}}
\put(1116.0,113.0){\rule[-0.500pt]{4.818pt}{1.000pt}}
\put(176.0,208.0){\rule[-0.500pt]{4.818pt}{1.000pt}}
\put(1116.0,208.0){\rule[-0.500pt]{4.818pt}{1.000pt}}
\put(176.0,303.0){\rule[-0.500pt]{4.818pt}{1.000pt}}
\put(154,303){\makebox(0,0)[r]{0.1}}
\put(1116.0,303.0){\rule[-0.500pt]{4.818pt}{1.000pt}}
\put(176.0,398.0){\rule[-0.500pt]{4.818pt}{1.000pt}}
\put(1116.0,398.0){\rule[-0.500pt]{4.818pt}{1.000pt}}
\put(176.0,493.0){\rule[-0.500pt]{4.818pt}{1.000pt}}
\put(154,493){\makebox(0,0)[r]{0.3}}
\put(1116.0,493.0){\rule[-0.500pt]{4.818pt}{1.000pt}}
\put(176.0,587.0){\rule[-0.500pt]{4.818pt}{1.000pt}}
\put(1116.0,587.0){\rule[-0.500pt]{4.818pt}{1.000pt}}
\put(176.0,682.0){\rule[-0.500pt]{4.818pt}{1.000pt}}
\put(154,682){\makebox(0,0)[r]{0.5}}
\put(1116.0,682.0){\rule[-0.500pt]{4.818pt}{1.000pt}}
\put(176.0,777.0){\rule[-0.500pt]{4.818pt}{1.000pt}}
\put(1116.0,777.0){\rule[-0.500pt]{4.818pt}{1.000pt}}
\put(176.0,872.0){\rule[-0.500pt]{4.818pt}{1.000pt}}
\put(154,872){\makebox(0,0)[r]{0.7}}
\put(1116.0,872.0){\rule[-0.500pt]{4.818pt}{1.000pt}}
\put(176.0,967.0){\rule[-0.500pt]{4.818pt}{1.000pt}}
\put(1116.0,967.0){\rule[-0.500pt]{4.818pt}{1.000pt}}
\put(176.0,113.0){\rule[-0.500pt]{1.000pt}{4.818pt}}
\put(176,68){\makebox(0,0){0}}
\put(176.0,947.0){\rule[-0.500pt]{1.000pt}{4.818pt}}
\put(272.0,113.0){\rule[-0.500pt]{1.000pt}{4.818pt}}
\put(272.0,947.0){\rule[-0.500pt]{1.000pt}{4.818pt}}
\put(368.0,113.0){\rule[-0.500pt]{1.000pt}{4.818pt}}
\put(368,68){\makebox(0,0){1}}
\put(368.0,947.0){\rule[-0.500pt]{1.000pt}{4.818pt}}
\put(464.0,113.0){\rule[-0.500pt]{1.000pt}{4.818pt}}
\put(464.0,947.0){\rule[-0.500pt]{1.000pt}{4.818pt}}
\put(560.0,113.0){\rule[-0.500pt]{1.000pt}{4.818pt}}
\put(560,68){\makebox(0,0){2}}
\put(560.0,947.0){\rule[-0.500pt]{1.000pt}{4.818pt}}
\put(656.0,113.0){\rule[-0.500pt]{1.000pt}{4.818pt}}
\put(656.0,947.0){\rule[-0.500pt]{1.000pt}{4.818pt}}
\put(752.0,113.0){\rule[-0.500pt]{1.000pt}{4.818pt}}
\put(752,68){\makebox(0,0){3}}
\put(752.0,947.0){\rule[-0.500pt]{1.000pt}{4.818pt}}
\put(848.0,113.0){\rule[-0.500pt]{1.000pt}{4.818pt}}
\put(848.0,947.0){\rule[-0.500pt]{1.000pt}{4.818pt}}
\put(944.0,113.0){\rule[-0.500pt]{1.000pt}{4.818pt}}
\put(944,68){\makebox(0,0){4}}
\put(944.0,947.0){\rule[-0.500pt]{1.000pt}{4.818pt}}
\put(1040.0,113.0){\rule[-0.500pt]{1.000pt}{4.818pt}}
\put(1040.0,947.0){\rule[-0.500pt]{1.000pt}{4.818pt}}
\put(1136.0,113.0){\rule[-0.500pt]{1.000pt}{4.818pt}}
\put(1136,68){\makebox(0,0){5}}
\put(1136.0,947.0){\rule[-0.500pt]{1.000pt}{4.818pt}}
\put(176.0,113.0){\rule[-0.500pt]{231.264pt}{1.000pt}}
\put(1136.0,113.0){\rule[-0.500pt]{1.000pt}{205.729pt}}
\put(176.0,967.0){\rule[-0.500pt]{231.264pt}{1.000pt}}
\put(656,-3){\makebox(0,0){$\Gamma t$}}
\put(700,800){\makebox(0,0){$\scriptstyle D^0(t)\rightarrow K^+\pi^-$}}
\put(700,760){\makebox(0,0){vs}}
\put(700,720){\makebox(0,0){$\scriptstyle \bar{D}^0(t)\rightarrow K^-\pi^+$}}
\put(700,330){\makebox(0,0){$\scriptstyle \bar{D}^0(t)\rightarrow K^+\pi^-$}}
\put(700,290){\makebox(0,0){vs}}
\put(700,250){\makebox(0,0){$\scriptstyle D^0(t)\rightarrow K^-\pi^+$}}

\put(176.0,113.0){\rule[-0.500pt]{1.000pt}{205.729pt}}
\put(176,208){\usebox{\plotpoint}}
\put(176,205.42){\rule{4.577pt}{1.000pt}}
\multiput(176.00,205.92)(9.500,-1.000){2}{\rule{2.289pt}{1.000pt}}
\put(234,204.42){\rule{4.577pt}{1.000pt}}
\multiput(234.00,204.92)(9.500,-1.000){2}{\rule{2.289pt}{1.000pt}}
\put(195.0,207.0){\rule[-0.500pt]{9.395pt}{1.000pt}}
\put(291,203.42){\rule{4.577pt}{1.000pt}}
\multiput(291.00,203.92)(9.500,-1.000){2}{\rule{2.289pt}{1.000pt}}
\put(253.0,206.0){\rule[-0.500pt]{9.154pt}{1.000pt}}
\put(330,202.42){\rule{4.577pt}{1.000pt}}
\multiput(330.00,202.92)(9.500,-1.000){2}{\rule{2.289pt}{1.000pt}}
\put(310.0,205.0){\rule[-0.500pt]{4.818pt}{1.000pt}}
\put(387,201.42){\rule{4.577pt}{1.000pt}}
\multiput(387.00,201.92)(9.500,-1.000){2}{\rule{2.289pt}{1.000pt}}
\put(349.0,204.0){\rule[-0.500pt]{9.154pt}{1.000pt}}
\put(426,200.42){\rule{4.577pt}{1.000pt}}
\multiput(426.00,200.92)(9.500,-1.000){2}{\rule{2.289pt}{1.000pt}}
\put(406.0,203.0){\rule[-0.500pt]{4.818pt}{1.000pt}}
\put(483,199.42){\rule{4.577pt}{1.000pt}}
\multiput(483.00,199.92)(9.500,-1.000){2}{\rule{2.289pt}{1.000pt}}
\put(445.0,202.0){\rule[-0.500pt]{9.154pt}{1.000pt}}
\put(522,198.42){\rule{4.577pt}{1.000pt}}
\multiput(522.00,198.92)(9.500,-1.000){2}{\rule{2.289pt}{1.000pt}}
\put(502.0,201.0){\rule[-0.500pt]{4.818pt}{1.000pt}}
\put(579,197.42){\rule{4.577pt}{1.000pt}}
\multiput(579.00,197.92)(9.500,-1.000){2}{\rule{2.289pt}{1.000pt}}
\put(541.0,200.0){\rule[-0.500pt]{9.154pt}{1.000pt}}
\put(618,196.42){\rule{4.577pt}{1.000pt}}
\multiput(618.00,196.92)(9.500,-1.000){2}{\rule{2.289pt}{1.000pt}}
\put(598.0,199.0){\rule[-0.500pt]{4.818pt}{1.000pt}}
\put(675,195.42){\rule{4.577pt}{1.000pt}}
\multiput(675.00,195.92)(9.500,-1.000){2}{\rule{2.289pt}{1.000pt}}
\put(637.0,198.0){\rule[-0.500pt]{9.154pt}{1.000pt}}
\put(714,194.42){\rule{4.577pt}{1.000pt}}
\multiput(714.00,194.92)(9.500,-1.000){2}{\rule{2.289pt}{1.000pt}}
\put(694.0,197.0){\rule[-0.500pt]{4.818pt}{1.000pt}}
\put(771,193.42){\rule{4.577pt}{1.000pt}}
\multiput(771.00,193.92)(9.500,-1.000){2}{\rule{2.289pt}{1.000pt}}
\put(733.0,196.0){\rule[-0.500pt]{9.154pt}{1.000pt}}
\put(810,192.42){\rule{4.577pt}{1.000pt}}
\multiput(810.00,192.92)(9.500,-1.000){2}{\rule{2.289pt}{1.000pt}}
\put(790.0,195.0){\rule[-0.500pt]{4.818pt}{1.000pt}}
\put(867,191.42){\rule{4.577pt}{1.000pt}}
\multiput(867.00,191.92)(9.500,-1.000){2}{\rule{2.289pt}{1.000pt}}
\put(829.0,194.0){\rule[-0.500pt]{9.154pt}{1.000pt}}
\put(906,190.42){\rule{4.577pt}{1.000pt}}
\multiput(906.00,190.92)(9.500,-1.000){2}{\rule{2.289pt}{1.000pt}}
\put(886.0,193.0){\rule[-0.500pt]{4.818pt}{1.000pt}}
\put(963,189.42){\rule{4.577pt}{1.000pt}}
\multiput(963.00,189.92)(9.500,-1.000){2}{\rule{2.289pt}{1.000pt}}
\put(925.0,192.0){\rule[-0.500pt]{9.154pt}{1.000pt}}
\put(1002,188.42){\rule{4.577pt}{1.000pt}}
\multiput(1002.00,188.92)(9.500,-1.000){2}{\rule{2.289pt}{1.000pt}}
\put(982.0,191.0){\rule[-0.500pt]{4.818pt}{1.000pt}}
\put(1059,187.42){\rule{4.577pt}{1.000pt}}
\multiput(1059.00,187.92)(9.500,-1.000){2}{\rule{2.289pt}{1.000pt}}
\put(1021.0,190.0){\rule[-0.500pt]{9.154pt}{1.000pt}}
\put(1098,186.42){\rule{4.577pt}{1.000pt}}
\multiput(1098.00,186.92)(9.500,-1.000){2}{\rule{2.289pt}{1.000pt}}
\put(1078.0,189.0){\rule[-0.500pt]{4.818pt}{1.000pt}}
\put(1117.0,188.0){\rule[-0.500pt]{4.577pt}{1.000pt}}
\put(176,208){\usebox{\plotpoint}}
\multiput(177.83,208.00)(0.495,1.473){30}{\rule{0.119pt}{3.197pt}}
\multiput(173.92,208.00)(19.000,49.364){2}{\rule{1.000pt}{1.599pt}}
\multiput(196.83,264.00)(0.495,1.473){30}{\rule{0.119pt}{3.197pt}}
\multiput(192.92,264.00)(19.000,49.364){2}{\rule{1.000pt}{1.599pt}}
\multiput(215.83,320.00)(0.495,1.372){32}{\rule{0.119pt}{3.000pt}}
\multiput(211.92,320.00)(20.000,48.773){2}{\rule{1.000pt}{1.500pt}}
\multiput(235.83,375.00)(0.495,1.393){30}{\rule{0.119pt}{3.039pt}}
\multiput(231.92,375.00)(19.000,46.691){2}{\rule{1.000pt}{1.520pt}}
\multiput(254.83,428.00)(0.495,1.339){30}{\rule{0.119pt}{2.934pt}}
\multiput(250.92,428.00)(19.000,44.910){2}{\rule{1.000pt}{1.467pt}}
\multiput(273.83,479.00)(0.495,1.285){30}{\rule{0.119pt}{2.829pt}}
\multiput(269.92,479.00)(19.000,43.128){2}{\rule{1.000pt}{1.414pt}}
\multiput(292.83,528.00)(0.495,1.177){30}{\rule{0.119pt}{2.618pt}}
\multiput(288.92,528.00)(19.000,39.565){2}{\rule{1.000pt}{1.309pt}}
\multiput(311.83,573.00)(0.495,1.040){32}{\rule{0.119pt}{2.350pt}}
\multiput(307.92,573.00)(20.000,37.122){2}{\rule{1.000pt}{1.175pt}}
\multiput(331.83,615.00)(0.495,1.015){30}{\rule{0.119pt}{2.303pt}}
\multiput(327.92,615.00)(19.000,34.221){2}{\rule{1.000pt}{1.151pt}}
\multiput(350.83,654.00)(0.495,0.907){30}{\rule{0.119pt}{2.092pt}}
\multiput(346.92,654.00)(19.000,30.658){2}{\rule{1.000pt}{1.046pt}}
\multiput(369.83,689.00)(0.495,0.827){30}{\rule{0.119pt}{1.934pt}}
\multiput(365.92,689.00)(19.000,27.985){2}{\rule{1.000pt}{0.967pt}}
\multiput(388.83,721.00)(0.495,0.719){30}{\rule{0.119pt}{1.724pt}}
\multiput(384.92,721.00)(19.000,24.422){2}{\rule{1.000pt}{0.862pt}}
\multiput(407.83,749.00)(0.495,0.606){32}{\rule{0.119pt}{1.500pt}}
\multiput(403.92,749.00)(20.000,21.887){2}{\rule{1.000pt}{0.750pt}}
\multiput(427.83,774.00)(0.495,0.557){30}{\rule{0.119pt}{1.408pt}}
\multiput(423.92,774.00)(19.000,19.078){2}{\rule{1.000pt}{0.704pt}}
\multiput(445.00,797.83)(0.476,0.495){30}{\rule{1.250pt}{0.119pt}}
\multiput(445.00,793.92)(16.406,19.000){2}{\rule{0.625pt}{1.000pt}}
\multiput(464.00,816.83)(0.567,0.494){24}{\rule{1.438pt}{0.119pt}}
\multiput(464.00,812.92)(16.016,16.000){2}{\rule{0.719pt}{1.000pt}}
\multiput(483.00,832.83)(0.700,0.492){18}{\rule{1.712pt}{0.118pt}}
\multiput(483.00,828.92)(15.448,13.000){2}{\rule{0.856pt}{1.000pt}}
\multiput(502.00,845.83)(0.878,0.489){14}{\rule{2.068pt}{0.118pt}}
\multiput(502.00,841.92)(15.707,11.000){2}{\rule{1.034pt}{1.000pt}}
\multiput(522.00,856.83)(1.158,0.481){8}{\rule{2.625pt}{0.116pt}}
\multiput(522.00,852.92)(13.552,8.000){2}{\rule{1.313pt}{1.000pt}}
\multiput(541.00,864.84)(1.606,0.462){4}{\rule{3.417pt}{0.111pt}}
\multiput(541.00,860.92)(11.909,6.000){2}{\rule{1.708pt}{1.000pt}}
\multiput(560.00,870.86)(2.188,0.424){2}{\rule{4.050pt}{0.102pt}}
\multiput(560.00,866.92)(10.594,5.000){2}{\rule{2.025pt}{1.000pt}}
\put(579,873.42){\rule{4.577pt}{1.000pt}}
\multiput(579.00,871.92)(9.500,3.000){2}{\rule{2.289pt}{1.000pt}}
\put(598,875.92){\rule{4.818pt}{1.000pt}}
\multiput(598.00,874.92)(10.000,2.000){2}{\rule{2.409pt}{1.000pt}}
\put(637,876.42){\rule{4.577pt}{1.000pt}}
\multiput(637.00,876.92)(9.500,-1.000){2}{\rule{2.289pt}{1.000pt}}
\put(656,874.92){\rule{4.577pt}{1.000pt}}
\multiput(656.00,875.92)(9.500,-2.000){2}{\rule{2.289pt}{1.000pt}}
\put(675,872.42){\rule{4.577pt}{1.000pt}}
\multiput(675.00,873.92)(9.500,-3.000){2}{\rule{2.289pt}{1.000pt}}
\put(694,868.92){\rule{4.818pt}{1.000pt}}
\multiput(694.00,870.92)(10.000,-4.000){2}{\rule{2.409pt}{1.000pt}}
\put(714,864.92){\rule{4.577pt}{1.000pt}}
\multiput(714.00,866.92)(9.500,-4.000){2}{\rule{2.289pt}{1.000pt}}
\multiput(733.00,862.71)(2.188,-0.424){2}{\rule{4.050pt}{0.102pt}}
\multiput(733.00,862.92)(10.594,-5.000){2}{\rule{2.025pt}{1.000pt}}
\multiput(752.00,857.71)(2.188,-0.424){2}{\rule{4.050pt}{0.102pt}}
\multiput(752.00,857.92)(10.594,-5.000){2}{\rule{2.025pt}{1.000pt}}
\multiput(771.00,852.69)(1.606,-0.462){4}{\rule{3.417pt}{0.111pt}}
\multiput(771.00,852.92)(11.909,-6.000){2}{\rule{1.708pt}{1.000pt}}
\multiput(790.00,846.69)(1.708,-0.462){4}{\rule{3.583pt}{0.111pt}}
\multiput(790.00,846.92)(12.563,-6.000){2}{\rule{1.792pt}{1.000pt}}
\multiput(810.00,840.69)(1.606,-0.462){4}{\rule{3.417pt}{0.111pt}}
\multiput(810.00,840.92)(11.909,-6.000){2}{\rule{1.708pt}{1.000pt}}
\multiput(829.00,834.69)(1.606,-0.462){4}{\rule{3.417pt}{0.111pt}}
\multiput(829.00,834.92)(11.909,-6.000){2}{\rule{1.708pt}{1.000pt}}
\multiput(848.00,828.69)(1.339,-0.475){6}{\rule{2.964pt}{0.114pt}}
\multiput(848.00,828.92)(12.847,-7.000){2}{\rule{1.482pt}{1.000pt}}
\multiput(867.00,821.69)(1.339,-0.475){6}{\rule{2.964pt}{0.114pt}}
\multiput(867.00,821.92)(12.847,-7.000){2}{\rule{1.482pt}{1.000pt}}
\multiput(886.00,814.69)(1.420,-0.475){6}{\rule{3.107pt}{0.114pt}}
\multiput(886.00,814.92)(13.551,-7.000){2}{\rule{1.554pt}{1.000pt}}
\multiput(906.00,807.69)(1.339,-0.475){6}{\rule{2.964pt}{0.114pt}}
\multiput(906.00,807.92)(12.847,-7.000){2}{\rule{1.482pt}{1.000pt}}
\multiput(925.00,800.69)(1.339,-0.475){6}{\rule{2.964pt}{0.114pt}}
\multiput(925.00,800.92)(12.847,-7.000){2}{\rule{1.482pt}{1.000pt}}
\multiput(944.00,793.69)(1.339,-0.475){6}{\rule{2.964pt}{0.114pt}}
\multiput(944.00,793.92)(12.847,-7.000){2}{\rule{1.482pt}{1.000pt}}
\multiput(963.00,786.69)(1.339,-0.475){6}{\rule{2.964pt}{0.114pt}}
\multiput(963.00,786.92)(12.847,-7.000){2}{\rule{1.482pt}{1.000pt}}
\multiput(982.00,779.69)(1.420,-0.475){6}{\rule{3.107pt}{0.114pt}}
\multiput(982.00,779.92)(13.551,-7.000){2}{\rule{1.554pt}{1.000pt}}
\multiput(1002.00,772.69)(1.339,-0.475){6}{\rule{2.964pt}{0.114pt}}
\multiput(1002.00,772.92)(12.847,-7.000){2}{\rule{1.482pt}{1.000pt}}
\multiput(1021.00,765.69)(1.339,-0.475){6}{\rule{2.964pt}{0.114pt}}
\multiput(1021.00,765.92)(12.847,-7.000){2}{\rule{1.482pt}{1.000pt}}
\multiput(1040.00,758.69)(1.339,-0.475){6}{\rule{2.964pt}{0.114pt}}
\multiput(1040.00,758.92)(12.847,-7.000){2}{\rule{1.482pt}{1.000pt}}
\multiput(1059.00,751.69)(1.339,-0.475){6}{\rule{2.964pt}{0.114pt}}
\multiput(1059.00,751.92)(12.847,-7.000){2}{\rule{1.482pt}{1.000pt}}
\multiput(1078.00,744.69)(1.420,-0.475){6}{\rule{3.107pt}{0.114pt}}
\multiput(1078.00,744.92)(13.551,-7.000){2}{\rule{1.554pt}{1.000pt}}
\multiput(1098.00,737.69)(1.339,-0.475){6}{\rule{2.964pt}{0.114pt}}
\multiput(1098.00,737.92)(12.847,-7.000){2}{\rule{1.482pt}{1.000pt}}
\multiput(1117.00,730.69)(1.606,-0.462){4}{\rule{3.417pt}{0.111pt}}
\multiput(1117.00,730.92)(11.909,-6.000){2}{\rule{1.708pt}{1.000pt}}
\put(618.0,879.0){\rule[-0.500pt]{4.577pt}{1.000pt}}
\end{picture}
\vspace{0.4cm}
\caption{Illustrative plot for changes of $CP$-violating asymmetries
with proper time $t$, where $x^{~}_D=0.05$, $y^{~}_D =0$, $\delta_{K\pi}=0$ 
and $\phi_D=\pi/2$ have been taken.}
\end{figure}

\section{Conclusion}

We have highlit some possible signals of $D^0$-$\bar{D}^0$ mixing and
$CP$ violation in neutral $D$-meson decays. Quantitatively, it remains 
difficult (even impossible) to give reliable predictions for most of 
such signals. Some progress can certainly be made in this topic if the future 
experiments are able to probe the $D^0$-$\bar{D}^0$ mixing rate
$r^{~}_D$ down to the $10^{-4}$ level and to search for $CP$-violating 
asymmetries down to the $10^{-3}$ level. The emergence of new physics
in the charm sector would offer a reward for all sophisticated experimental 
efforts which are underway today.

\vspace{0.3cm}
\begin{flushleft}
{\Large\bf Acknowledgments}
\end{flushleft}

I would like to thank S. Pakvasa for inviting me to participate in 
this nice conference. I am grateful to A.I. Sanda for his encouragement
and support, which make my participation realizable.
This work was supported by the Japan Society for the Promotion of Science.

\end{document}